\documentclass[12pt]{iopart}
\usepackage{graphicx}
\DeclareGraphicsExtensions{.pdf,.png,.jpg,.eps}
\usepackage{iopams}  

\begin{document}
\title[Ultracold few fermionic atoms in needle-shaped double wells]{Ultracold few fermionic atoms in needle-shaped
double wells: spin chains and resonating spin clusters from microscopic Hamiltonians emulated via 
antiferromagnetic Heisenberg and $t$-$J$ models}

\author{Constantine Yannouleas}
\ead{Constantine.Yannouleas@physics.gatech.edu}
\address{School of Physics, Georgia Institute of Technology,
             Atlanta, Georgia 30332-0430, USA}
\author{Benedikt B. Brandt}
\ead{benbra@gatech.edu}
\address{School of Physics, Georgia Institute of Technology,
             Atlanta, Georgia 30332-0430, USA}
\author{Uzi Landman}
\ead{Uzi.Landman@physics.gatech.edu}
\address{School of Physics, Georgia Institute of Technology,
             Atlanta, Georgia 30332-0430, USA}
\vspace{12pt}
\begin{indented}
\item[]19 May 2016
\end{indented}

\begin{abstract}
Advances with trapped ultracold atoms intensified interest in simulating complex physical phenomena, including 
quantum magnetism and transitions from itinerant to non-itinerant behavior. Here we show formation of 
antiferromagnetic ground states of few ultracold fermionic atoms in single and double well (DW) traps, through 
microscopic Hamiltonian exact diagonalization for two DW arrangements: (i) two linearly oriented one-dimensional, 
1D, wells, and (ii) two coupled parallel wells, forming a trap of two-dimensional, 2D, nature. The spectra and 
spin-resolved conditional probabilities reveal for both cases, under strong repulsion, atomic spatial localization 
at extemporaneously created sites, forming quantum molecular magnetic structures with non-itinerant character.
These findings usher future theoretical and experimental explorations into the highly-correlated behavior of 
ultracold strongly-repelling fermionic atoms in higher dimensions, beyond the fermionization physics that is 
strictly applicable only in the 1D case. The results for four atoms are well described with finite Heisenberg 
spin-chain and cluster models. The numerical simulations of three fermionic atoms in symmetric double wells reveal 
the emergent appearance of coupled resonating 2D Heisenberg clusters, whose emulation requires the use of a 
$t$-$J$-like model, akin to that used in investigations of high T$_c$ superconductivity. The highly entangled 
states discovered in the microscopic and model calculations of controllably detuned, asymmetric, double wells 
suggest three-cold-atom DW quantum computing qubits.
\end{abstract}
~~~~~~\\

%
%
Published: New J. Phys. {\bf 18}, 073018 (2016)
%
%

~~~~~~\\
~~~~~~\\
~~~~~~\\

\section{Introduction}
\label{intro}

The unparalleled experimental advances and control achieved in the field of ultracold atoms have rekindled an
intense interest in emulating magnetic behavior using ultracold atoms in optical traps \cite{cira04,grei11}.
Quantum magnetism and spintronics in both extended \cite{zene511,ande555,auerbookk,fazebookk,zuti04} and 
finite-size \cite{hend933,bade06,haas088,yann07,yann09} systems have a long history. Recently antiferromagnetism 
(AFM) without the assistance of an external periodic ordering potential has been demonstrated experimentally for 
$N=3$ and $N=4$ ultracold $^6$Li atoms confined in a single-well one-dimensional optical trap \cite{murm15.2}.
Moreover, progress aiming at bottom-up approaches to fermionic many-body systems, addressing entanglement, 
quantum information, and quantum magnetism in particular, is predicated on experimental developments of which the 
recently created double-well (DW) ultracold atom traps \cite{kauf14,murm15.1,kauf15} are the first steps.

\begin{figure}[t]
\centering\includegraphics[width=11cm]{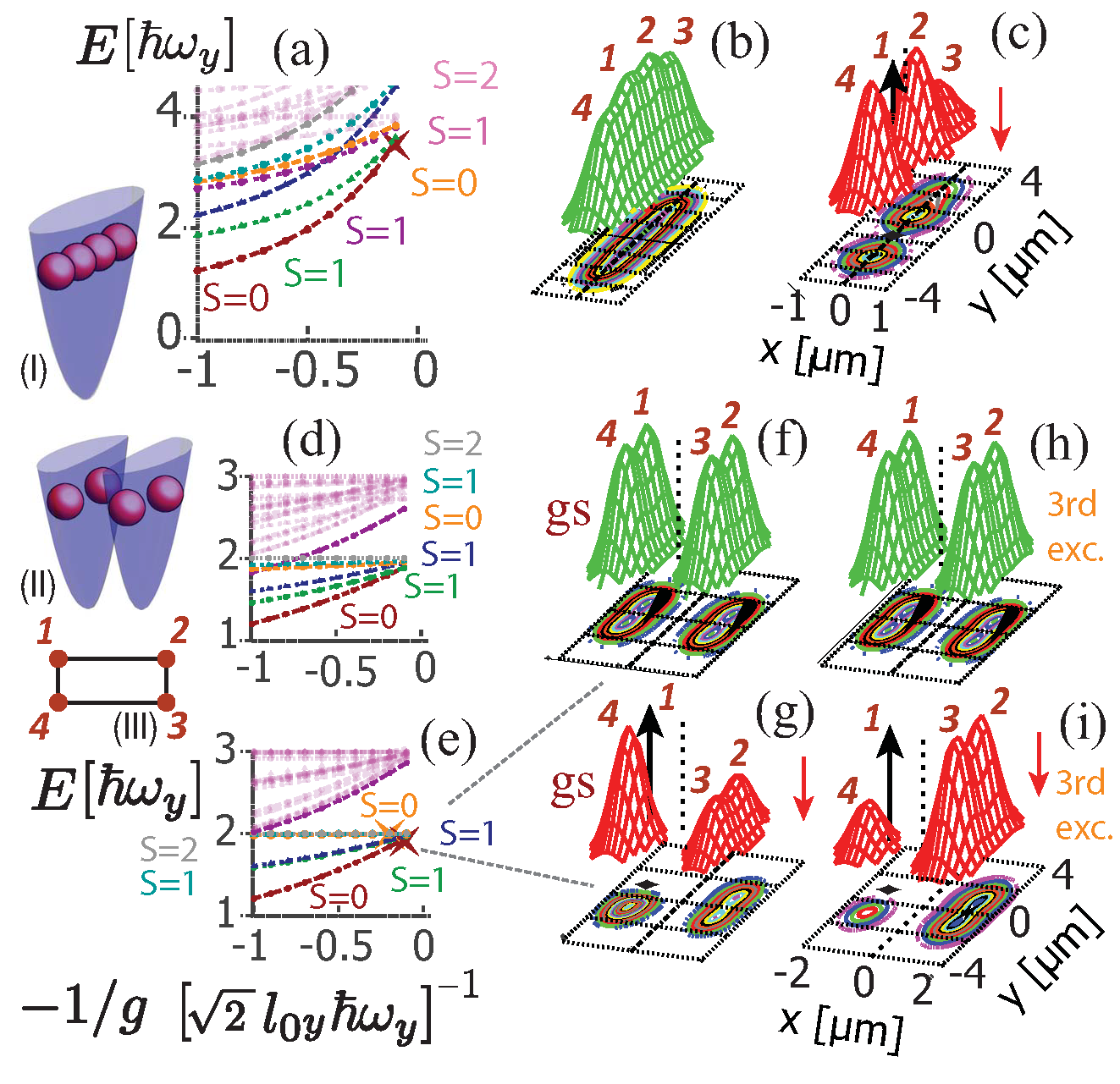}
\caption{Energy versus $-1/g$ spectra, SPDs (green surfaces), and spin-resolved CPDs 
(red surfaces) of $N=4$ strongly-repelling $^6$Li 
atoms in a double-well confinement with a {\it parallel\/} arrangement of the two 1D traps, as a function of the
interwell separation $d$ and/or interwell barrier $V_b$. Schematic (I) shows a SW in the $y$-direction, and (II) 
displays a symmetric DWPA, where interwell tunneling in the $x$-direction occurs along the entire $y$-range of the 
trap, conferring 2D aspects to the trap.  Insert (III) shows the sites in the 2D Heisenberg-ring spin model. 
(a,b,c) $d=0$ (single well). (d) $d=2.5$ $\mu$m and $V_b=6.08$ kHz (lower barrier). (e,f,g,h,i) $d=2.5$ $\mu$m and 
$V_b=11.14$ kHz (high barrier). In all cases, the confinement frequencies of the 1D traps are 
$\hbar \omega_x=6.6$ kHz and $\hbar \omega_y=1$ kHz. The SPDs and CPDs in (b,c) and (f,g) correspond to the 
$S=0$, $S_z=0$ CI ground states [gs, brown curves in the associated spectra 
(a) and (e)] at the point (marked by a star) 
$-1/g=-0.1 (\sqrt{2}l_{0y} \hbar \omega_y )^{-1}$. The SPD and CPD in (h,i) correspond to the $S=0$, $S_z=0$ CI 
excited state [orange curve in the associated spectrum (e)]. $g$ here is the 1D contact-interaction strength 
along the $y$ direction \cite{yann15}. All three CPDs display the distributions of the two down spins when the fixed 
spin-up fermion (see the black arrow) is placed at ${\bf r}_0=(0,-0.8$ $\mu$m) in (c), 
${\bf r}_0=(-1.3$ $\mu$m, $1$ $\mu$m) in (g), and ${\bf r}_0=(-1.28$ $\mu$m, $0.94$ $\mu$m) in (i).
$l_{0y(x)}=[\hbar/(M \omega_{y(x)})]^{1/2}$ is the harmonic-oscillator length; $M=9.99 \times 10^{-27}$ kg is the
mass of $^6$Li. The zero of energy in the spectra corresponds to the ground-state total energy of the corresponding 
non-interacting system, that is, to $4 \hbar \omega_y+ 2 \hbar \omega_x = 17.20$ $\hbar \omega_y$ in (a), 13.45 
$\hbar \omega_y$ in (d), and 14.91 $\hbar \omega_y$ in (e).
} 
\label{specpa}
\end{figure}

To date the theoretical studies of magnetism of a few ultracold atoms have mainly addressed 
\cite{deur08,deur14,zinn14,zinn15,bruu15,ho14} strictly one-dimensional systems trapped within a single well
(see, however, Refs.\ \cite{zinn14,zinn15} for double-well configurations),  
where the fermionization theory \cite{gira10,gira07,guan09} (applicable to 1D systems in the limit of infinite
strength of the contact interaction) can assist in inventing analytic forms for the correlated many-body wave 
functions. However, the recently demonstrated ability to create needle-shaped double well traps \cite{murm15.1}, and 
the anticipated near-future further development of small arrays of such needle-like quasi-1D traps in a parallel
arrangement (PA, see schematics in Fig.\ \ref{specpa}), whose corresponding physics incorporates certain 
two-dimensional (2D) aspects \cite{yann15}, enjoin the development of additional conceptual and computational 
theoretical methodologies.

We remark that the physics of ultracold atoms in 1D and 3D single traps has been investigated also away from the
fermionization point using a Lippman-Schwinger equation approach, see Refs.\ \cite{blum12,blum13} and \cite{blum15}, 
respectively. Similarly, states of ultracold fermions in a single strictly-1D trap, away from the fermionization limit,
using an exact diagonalization of the many-body Hamiltonian have been reported \cite{ront14}. 

In this paper, using large-scale configuration-interaction (CI) calculations
as means for exact diagonalization of the microscopic Hamiltonian,
we report that for $N = 4$ (even number) strongly interacting ultracold fermions in a double-well trap with parallel
arrangement [DWPA \cite{yann15}; see Figs.\ \ref{specpa}(II,III)] the many-body problem can be reduced 
to that of a 2D rectangular AFM Heisenberg ring. The associated mapping between the many-body wave function and the 
spin eigenfunctions \cite{pauncz} for $N=3$ and $N=4$ electrons confined in single and double semiconductor quantum 
dots has been predicted in previous studies \cite{yann07,yann09} to occur through the formation of quantum 
molecular structures in the 
regime of strong long-range Coulombic repulsion. Such molecular structures are usually referred to as Wigner 
molecules (WMs) \cite{yann07.2}. For $N = 3$ (odd number) ultracold fermions, few-body quantum magnetism requires 
introduction of a more complex $t$-$J$-type \cite{auerbookk,fazebookk} model; here the $t$-$J$ model consists of 
two coupled and resonating triangular 2D Heisenberg clusters. In all cases, we find AFM ordered ground states.

We remark that the emergence of a resonance associated with the symmetrization of the many-body wave function in
two-center/three-electron bonded systems is well known \cite{hibe94,bick98,berr16} in theoretical chemistry and
in particular in the valence-bond treatment of the three-electron bond which controls the formation of molecules like 
He$^+_2$, NO, and F$_2^-$. The concept of the three-electon resonant bond and its significance were introduced
in 1931 in a seminal paper by Linus Pauling \cite{paul31,paulbook}.

The emergence of the simple, as well as the resonating, Heisenberg clusters is a consequence of the spatial 
localization of the strongly-interacting, highly correlated fermionic atoms and the formation \cite{yann15} of 
quantum 1D and 2D molecule-like structures, referred to as ultracold Wigner molecules (UCWMs). 
The name of Wigner is used here in the context of ultracold atoms in order to emphasize the universal aspects that 
are present in the few-fermion molecular structures irrespective of the nature of the repelling two-body interaction,
i.e., contact versus long-range Coulomb. In this way the concept of UCWM extends and incorporates \cite{roma04} the 
fermionization physics \cite{gira10,gira07,guan09} beyond the restricted 1D case. 

We note that due to the quantum character \cite{yann07.2} of the WMs, the 
spatial localization of fermions is not necessarily pointlike as in the classical electronic Wigner crystal 
\cite{wign34,wign38,vignbook}. However, depending on the strength of the Coulomb repulsion compared to the quantal 
kinetic energy, the regime of high-degree localization can be reached also in Wigner molecules formed by electrons 
confined in quantum dots \cite{yann07.2,yann99,yann00}. In contrast, fermions interacting via a a repulsive contact 
interaction cannot attain a similar degree of strong spatial localization; as a result, the UCWMs retain their full 
quantal character even in the limit of infinite repulsion.    

The resonant coupling of magnetic configurations through the tunneling of electrons between occupied and empty
sites has long been studied. Two well-known relevant fields are: (i) the socalled direct exchange
mechanism \cite{zene511,ande555} (related to ferromagnetism in the mixed-valency manganites of perovskite structure),
and (ii) the $t$-$J$ model \cite{auerbookk,dago944} which modifies (away from the half filling) the antiferromagnetic
Heisenberg Hamiltonian associated with the Mott insulator at half-filling. The
$t$-$J$ model has attracted much attention, because it has been proposed for explaining the high-T$_c$
superconductivity arising in the case of underdoped insulators \cite{dago944}. Due to the antiferromagnetic aspect, 
our resonating model Hamiltonian for $N=3$ fermions (see Section \ref{3ftj} below) represents a finite variant of the
$t$-$J$ model. The emergence of the $t$-$J$ model in this work suggests future investigations of fundamental aspects 
associated with the physics of high-T$_c$ superconductivity via studies utilizing the ability to prepare and measure 
trapped ultracold fermionic atom systems with precise control over the number of atoms and the strength of interatomic 
interactions.

\begin{figure}[t]
\centering\includegraphics[width=8cm]{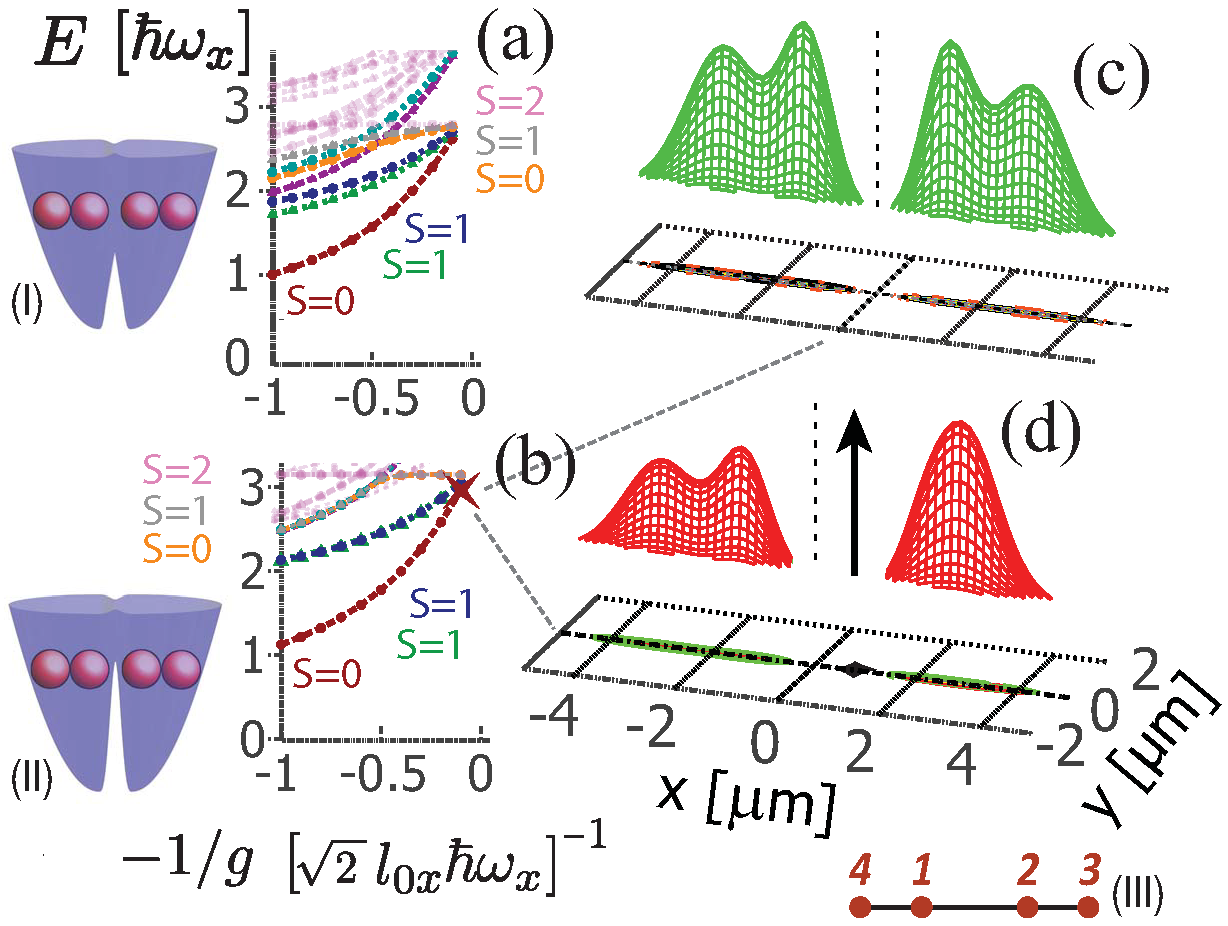}
\caption{Spectra, densities and spin-resolved CPDs of 4 fermionic atoms in linear double-well confinements.
Results are shown for two values of $V_b$; see schematics (I) and (II) with $V_b$ larger in (II). 
In the DWLI system, atomic motions in the wells and the interwell 1D tunneling occur along the $x$-axis.
Insert (III) shows the sites in the 1D Heisenberg-chain spin model.  
(a) $d=2.5$ $\mu$m and $V_b=2.3$ kHz (lower barrier). 
(b,c,d) $d=2.5$ $\mu$m and $V_b=8.5$ kHz (high barrier). In all cases, the confinement frequencies of the 1D traps 
are $\hbar \omega_x=1$ kHz and $\hbar \omega_y=100$ kHz. The SPD and CPD
in (c,d) correspond to the $S=0$, $S_z=0$ CI ground state [brown curve in the associated spectrum (b)] at the point 
(marked by a star) $-1/g=-0.1 (\sqrt{2}l_{0x} \hbar \omega_x )^{-1}$; 
$g$ here is the 1D contact-interaction strength along the $x$ direction \cite{yann15}. 
The CPD in (d) displays the distribution of the two down spins when the fixed spin-up fermion (see the black arrow) 
is placed at ${\bf r}_0=(0,$0.8 $\mu$m).  The zero of energy in the spectra corresponds to the ground-state total 
energy of the corresponding non-interacting system, that is, to 202.86 $\hbar \omega_x$ in (a) and 
203.52 $\hbar \omega_x$ in (b). 
The difference in the non-interacting energies between (a) and (b) is due to the different interwell barrier.
For two wells at infinite separation, the total energy for 4 non-interacting fermions is equal
to $2 \hbar \omega_x + 2 \hbar \omega_y = 202$ $\hbar \omega_x$.
}
\label{specli}
\end{figure}

We complement our investigations by further highlighting the differences arising from the different geometries of 
the traps in both the parallel and linear (LI) arrangement; in analogy to the DWPA designation defined above, 
a double-well trap with linear arrangement of the two needle-like wells will be denoted as DWLI; see schematics in
Figs.\ \ref{specli}(I,II).
Our theoretical predictions can be directly confirmed using the recently developed experimental techniques.
We stress again that the regime of ultracold Wigner-molecule (UCWM) formation and of the associated
simple-Heisenberg-chain and $t$-$J$ resonating-spin-chains magnetism appears for strong interparticle
interactions and contrasts sharply with the regime of itinerant magnetism \cite{heis11,kett09},
which appears for weaker interactions. The microscopic treatment of itinerant magnetism (weaker interactions) can 
be handled within mean-field approaches (e.g. Hartree-Fock), whereas the regime of spin chains (strong repulsion) 
considered here entails conservation of the total spin and requires more sophisticated approaches like the full CI.

Finally, we note that the related three-electron system in semiconductor double quantum dots has recently
attracted a major attention in conjunction with the fabrication and implementation of pulse-gated fast hybrid qubits
for solid-state-based quantum computing \cite{copp14.1,copp14.2}. These advances and the fascinating physics of 
double-well-trapped three ultracold fermionic atoms that we uncover, and in particular the high degree of 
entanglement predicted by us for strong interatomic repulsion (see Sections \ref{4ferm}, \ref{3ferm}, and \ref{enta}) 
and the very slow decoherence in such traps, suggest future exploration of this system as a robust ulracold 3-atom DW 
qubit.

Before leaving the Introduction, we wish to clarify that the term antiferromagnetic is used by us to
characterize finite systems having a ground state with the minimum possible value of the total spin, i.e.,
$S=1/2$ for $N=3$ fermions and $S=0$ for $N=4$ fermions. 

The plan of the paper is as follows:

A statement of the many-body Hamiltonian, including a description of the double-well employed by us, is
given in Section \ref{mbh}. 

In Section \ref{4ferm}, we describe investigations concerning 
four ultracold $^6$Li atoms in double-well traps with both
parallel and linear arrangement of the individual needle-like wells. A comparison with the case of four fermions in a 
quasi-1D single well is also included in order to appreciate the rich additional magnetic behaviors associated with a
double well. Section \ref{4ferm} is divided into two subsections, with Sec.\ \ref{4fci} describing results of purely 
microscopic CI calculations, and Sec.\ \ref{4fhh} establishing the mapping onto the Heisenberg 4-fermion 
phenomenological Hamiltonian.

Section \ref{3ferm} presents our studies concerning the case of three ultracold $^6$Li atoms in DWPA and DWLI traps, 
as well as the comparison with the corresponding case of a single well. Sec.\ \ref{3fci} describes CI results for
both the symmetric and tilted cases; for the tilted case, this section establishes the mapping onto a 3-fermion
Heisenberg model. Going beyond the Heisenberg model, Sec.\ \ref{3ftj} introduces the $t-J$ model and establishes, in
analogy with the CI results, its validity for describing the case of symmetric double wells.

Section \ref{enta} describes the entanglement properties of the CI many-body wave functions.

The appendices provide detailed information concerning the mathematical formalism associated with the spin 
eigenfunctions and the finite Heisenberg and $t-J$ models. In particular,
 
\ref{speig} provides a brief description of the branching diagram that describes the multiplicities (number of 
degenerate spin states) of a given total spin $S$. 
This Appendix also presents in the Ising basis the general formulas that describe a
spin eigenfunction (i) with $S=0$ and $S_z=0$ for four fermions and (ii) with $S=1/2$ and $S_z=1/2$ for
three fermions. These general formulas incorporate in a compact form both the orthogonal basis of spin functions that 
spans the spin space for a given $S$, as well as any linear superposition of them. In addition, \ref{speig} describes
the process of mapping the many-body CI wave functions onto these spin eigenfunctions.

\ref{ab} discusses the mathematics of the Heisenberg model for 4 localized fermions in a ring-like rectangular
configuration (DWPA case), while \ref{ac} discusses the corresponding case for an open chain arrangement
(DWLI case).

\ref{heis3} discusses the mathematics of the Heisenberg model for 3 localized fermions in a triangular
(DWPA case) and linear (DWLI case) configuration, both associated with tilted wells. 

Finally, \ref{tj3} discusses the mathematics of the more general $t-J$ model for 3 localized fermions in the case of
a double trap with symmetric wells.  
 
\begin{figure}[t]
\centering\includegraphics[width=7cm]{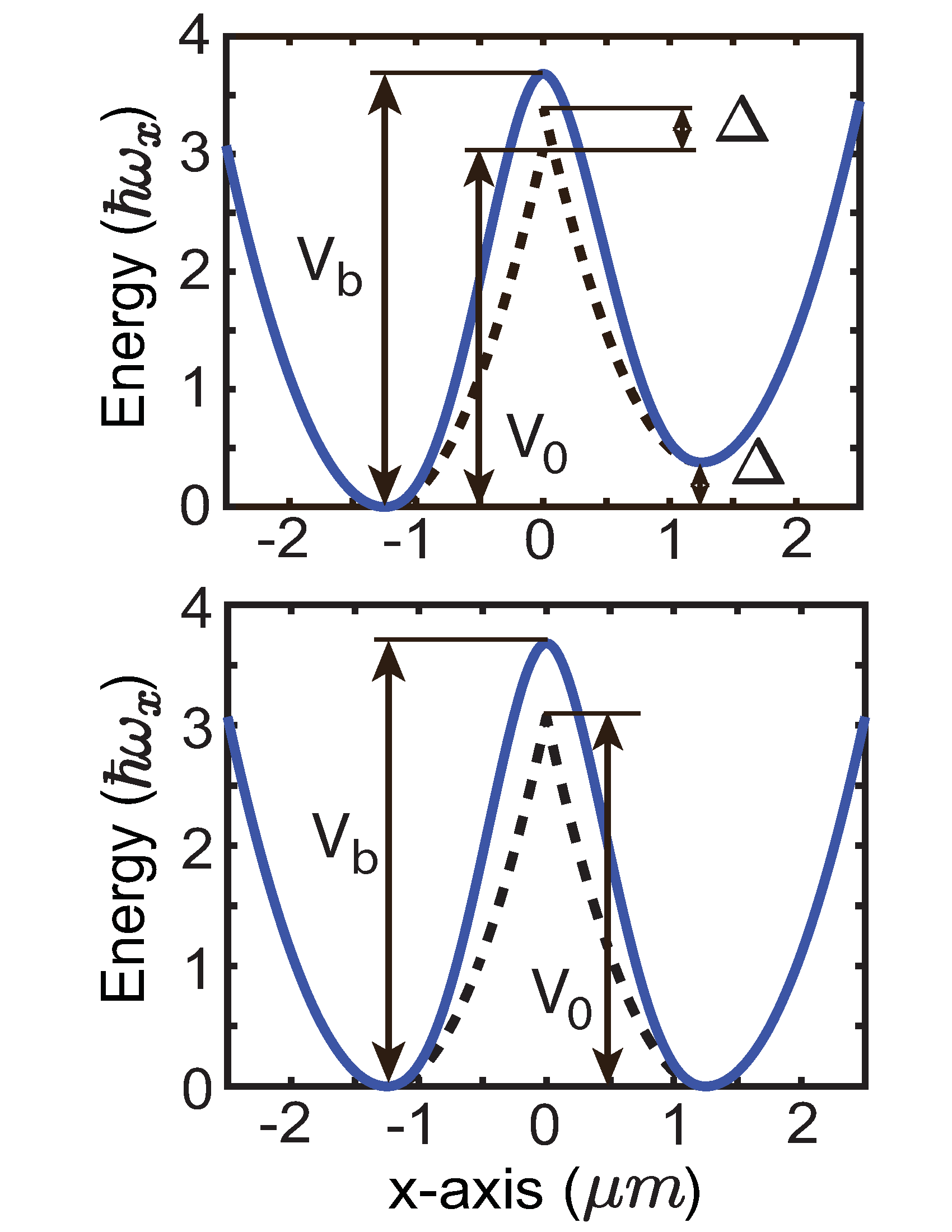}
\caption{TCO trapping potentials in the $x$ direction illustrating the smooth neck. (a) Tilted double well.
(b) Symmetric double well. The parameters correspond to the cases for three 
$^6$Li atoms in Figs.\ \ref{specn3}(f,j) below.
$\hbar \omega_x=6.6$ kHz, $V_b=24.30$ kHz, $d=2.5$ $\mu$m. $\Delta=2.5$ $\hbar \omega_y$ ($\hbar \omega_y=1$ kHz) 
in (a) and $\Delta=0$ in (b). $V_0$ denotes the interwell barrier (from the left side) for the pure 
two-parabola confinement without a smooth neck (dashed curve). When $\Delta \neq 0$, the dashed curve 
is not continuous at $x=0$; this is corrected with the consideration of the smooth neck.   
}
\label{tcopot}
\end{figure}

\section{Many-Body Hamiltonian}
\label{mbh}

A DWLI trap can be treated as a strictly-1D problem along a single direction. The DWPA trap which consists of two 
parallel needle-like wells, however, cannot be treated solely along one direction (e.g., the $x$-direction). Instead 
it requires consideration of the $y$ coordinate as well. To treat both cases in a unified way, we consider a 
many-body Hamiltonian for $N$ fermions of the form 
\begin{equation}
{\cal H}_{\rm MB} ({\bf r}_i,{\bf r}_j) =\sum_{i=1}^{N} H(i) +
\sum_{i=1}^{N} \sum_{j>i}^{N} g_{xy} \delta({\bf r}_i-{\bf r}_j)~,
\label{mbhd}
\end{equation}
where ${\bf r}_i-{\bf r}_j$ denotes the relative vector distance between the $i$ and $j$ fermions (e.g., $^6$Li 
atoms). This Hamiltonian is the sum of a single-particle part $H(i)$, which guarantees the needle-like shape of the
individual wells, and the two-particle contact interaction.

The external confining potential [in $H(i)$] that models a double well (DW) is based on a two-center-oscillator 
(TCO) model \cite{yann09,yann15} exhibiting a variable smooth neck along the $x$-direction. Along the $x$ direction, 
this TCO model allows for an independent variation of both the separation $d$ and of the barrier height $V_b$ between 
the two wells; see Fig.\ \ref{tcopot}. Along the $y$-direction, the confinement consists of that of a single 
harmonic oscillator. The values of the frequencies $\hbar \omega_{x1}$ (left well),  $\hbar \omega_{x2}$ (right well) 
and $\hbar \omega_y$ that confine the two wells along the $x$ and $y$ directions, respectively, are also allowed to 
vary independently; here we choose $\hbar \omega_{x1}=\hbar \omega_{x2}=\hbar \omega_x$. The needle-like shape of each
individual trap is enforced by assuming that $\hbar \omega_x << \hbar \omega_y$ (DWLI case) or 
$\hbar \omega_x >> \hbar \omega_y$ (DWPA case). The TCO further allows consideration of a tilt $\Delta$ between the 
left and right wells. Fig.\ \ref{tcopot} illustrates the TCO confining potentials in the $x$ direction used in Fig.\ 
\ref{specn3} below for the study of $N=3$ fermions in tilted and symmetric double wells. 

As we mentioned in the Introduction, we use the CI method for determining the solutions of the many-body problem 
specified by the Hamiltonian (\ref{mbhd}). The CI method expresses the fermionic many-body wave function as a 
supperposition of Slater determinants, and it is well known in quantum chemistry and in few-body physics of
electrons; for a basic description of the CI method, see Ref.\ \cite{szabobook}. Thus a detailed description of the 
CI method will not be repeated here. Specific adaptations by us of this method to a few electrons in 2D semiconductor 
quantum dots and rotating bosons in the lowest Landau level have been reported in Refs.\ \cite{yann09,yann07.2} and 
\cite{yann07.3}, respectively. An earlier application by us of this method to the case of $N=2$ trapped ultracold 
fermions was reported in Ref.\ \cite{yann15}. The reader can find an expanded description of the CI method in
the literature mentioned above. 

Convergence in the CI calculations is reached through the use of a basis of up to 
eighty TCO single-particle states as needed. Note that the TCO single-particle states automatically adjust to the 
separation $d$ as it varies from the limit of the unified atom $d=0$ to that of the two fully separated traps (for 
sufficiently large $d$). We verified that for $\omega_y/\omega_x=100$ (strictly-1D single trap), our CI 
calculations agree with the results of Table 2 of Ref.\ \cite{ront14}. 

The matrix elements of ${\cal H}_{\rm MB}$ between the CI determinants are calculated using the Slater rules.
An important ingredient in this respect are the two-body matrix elements of the contact interaction
\begin{equation}
g_{xy} \int_{-\infty}^\infty \int_{-\infty}^\infty 
d{\bf r}_1 d{\bf r}_2 \varphi^*_i({\bf r}_1) \varphi^*_j({\bf r}_2) \delta({\bf r}_1 - {\bf r}_2)
\varphi_k({\bf r}_1) \varphi_l({\bf r}_2),
\label{2ddelta}
\end{equation}
in the basis (of dimension $K$) formed out of the single-particle (space) eigenstates $\varphi_i({\bf r})$ of the TCO 
Hamiltonian. 

Because the individual wells remain needle-like in all of our calculations here, the $s$-wave scattering between two
ultracold fermions takes place primarily along a single dimension, either the $y$-dimension or the $x$-dimension.
As a result, the parameter $g_{xy}$ in front of the $\delta({\bf r}_i-{\bf r}_j)$ function in Eq.\ (\ref{mbhd})
does not reflect the physical process of two-dimensional $s$-scattering. Rather it is an auxiliary theoretical 
parameter that allows us to treat the DWPA and DWLI traps on an equal footing. In particular, the actual 1D 
interparticle interaction strengths, $g$, are related to $g_{xy}$ as follows
\begin{equation}
g = g_{xy} \int_{-\infty}^\infty du [W(u)]^4,
\label{g1d}
\end{equation}
where $u$ is a dummy variable and $W$ is the lowest-in-energy single-particle state in the $y$ $(x)$ direction
for the LI (PA) trap configurations, respectively.

Note that the 1D strength $g$ relates to the 3D $s$-scattering length $a_{3D}$ via the relation \cite{olsh98}, 
\begin{equation}
g=\frac{ 2 \hbar^2 a_{3D} } {\mu l_\perp^2} \frac{1}{ 1-1.4603 a_{3D}/l_\perp },
\label{ga3d}
\end{equation}
precisely as is done in the experimental studies of Ref.\ \cite{murm15.1}; $\mu=M/2$ is the relative mass and
$l_\perp$ is the harmonic oscillator strength in the direction perpendicular to the needle.

For the CI calculations in this paper, we assume that the total-spin projection $S_z=0$  for 4 fermions or $S_z=1/2$ 
for 3 fermions. This suffices to provide the full energy spectrum, as long as the many-body Hamiltonian does not 
depend on $S_z$. Naturally, the many-body wave functions characterized by a given total spin $S$ are different for
the different projections $S_z > 0$. For lack of space, we will not consider here many-body wave functions with 
$S_z \neq 0$ for four fermions or with $S_z \neq 1/2$ for three fermions. For an earlier study of such wave functions 
in the case of four electrons in a double quantum dot, see Ref.\ \cite{yann09}.     

\section{Four fermionic ultracold atoms in a double-well trap}
\label{4ferm}

\subsection{Four fermionic ultracold atoms: CI results}
\label{4fci}

We treat here three different types of 
traps: a SW quasi-1D trap [see Fig.\ \ref{specpa}(I)], a DWPA trap [parallel arrangement, Fig.\ \ref{specpa}(II)], 
and a DWLI trap [linear arrangement, Fig.\ \ref{specli}(I)]. 
We start with the four-atom double-well systems, followed by a comparison with the single-well trap
(end of Secs.\ \ref{4fci} and \ref{4fhh}), which is used as a reference point to allow for a deeper appreciation 
of the richness of magnetic behaviors introduced by the double-well geometries.

Figs.\ \ref{specpa} and \ref{specli} illustrate the evolution of the spectra of $N=4$ $^6$Li atoms for the DWPA and
DWLI cases, respectively, as a function of the separateness of (or alternatively the strength of tunneling between) 
the two 1D wells (resulting from both the effect of 
separation in distance, $d$, and the height of the interwell barrier $V_b$). The limiting case of the single-well 
(``united atom'') quasi-1D trap (at $d=0$) is displayed in Fig.\ \ref{specpa}(a). The opposite limit of 
strongly separated wells is displayed in Fig.\ \ref{specpa}(e) and Fig.\ \ref{specli}(b), respectively. 
All spectra are shown in the range $-1/(\sqrt{2}l_{0y(x)} \hbar \omega_{y(x)} ) \leq -1/g \leq 0$, 
which covers the regime of strong interparticle contact repulsion.
A salient common feature of all five energy spectra in Figs.\ \ref{specpa} and \ref{specli} is the emergence of a 
separate band formed by six low-energy states as the interaction strength approaches
infinity (i.e., as $-1/g \rightarrow -0$); all six states become degenerate at $-1/g=0$. Qualitative differences
between these spectra amount only in the extent of the spreading out of the six curves; the most spread out case (with 
six clearly distinct lines) arises for the single well [Fig.\ \ref{specpa}(a)], whereas the strongly separated cases 
display a characteristic 1-2-3 degeneracy in the whole range $-1 \leq -1/g \leq 0$ [Fig.\ \ref{specpa}(e) and Fig.\ 
\ref{specli}(b)]. The tendency towards the regrouping of the energy curves according to the 1-2-3 degeneracy pattern 
is also visible in the intermediate cases [Fig.\ \ref{specpa}(d) and Fig.\ \ref{specli}(a)]. 

\begin{figure}[t]
\centering\includegraphics[width=7.8cm]{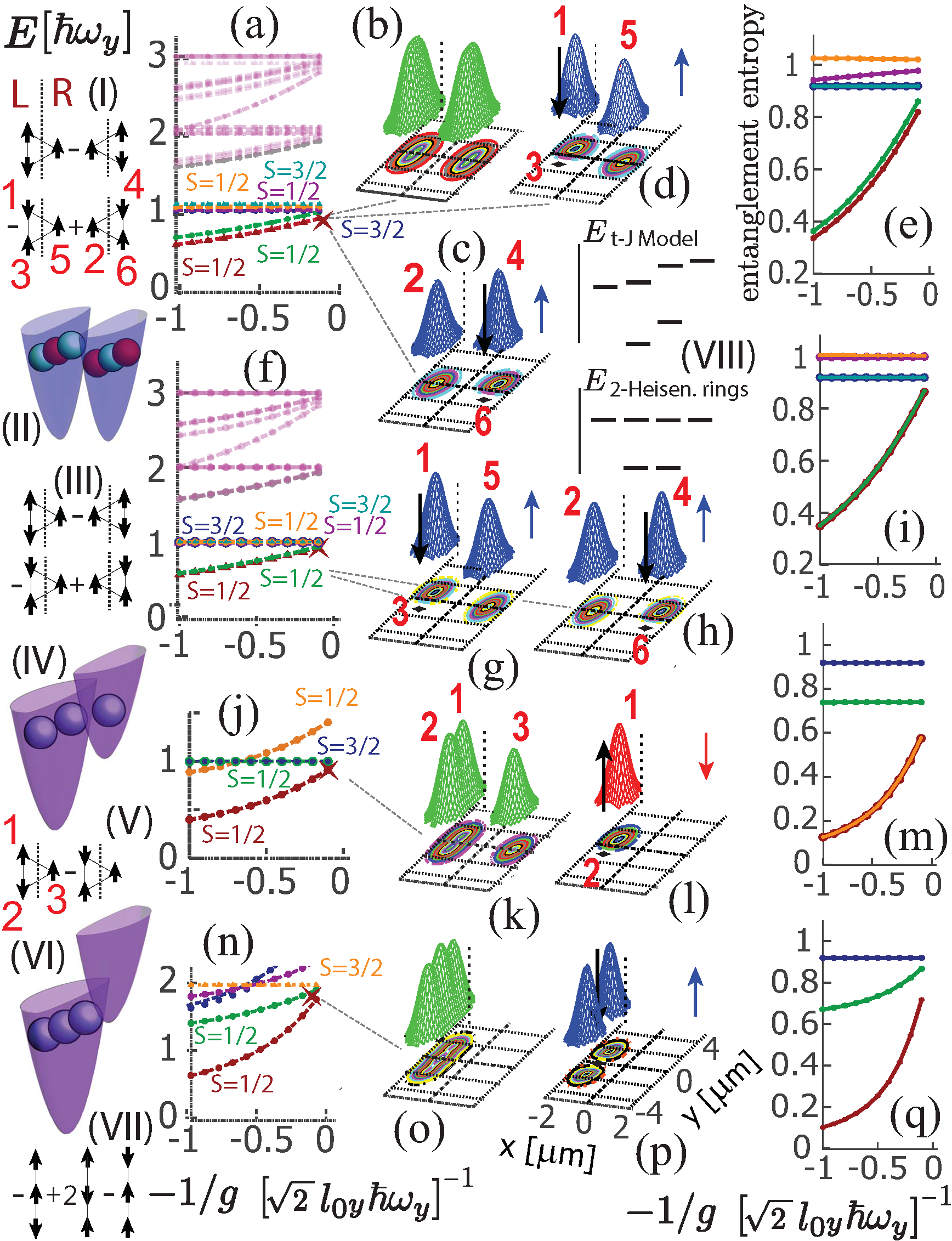}
\caption{Energy versus $-1/g$ spectra, SPDs (green surfaces), and spin-resolved CPDs 
of $N=3$ strongly-repelling $^6$Li 
atoms in a double-well confinement with a {\it parallel\/} arrangement (DWPA) of the two 1D traps at a given 
interwell separation $d=2.5$ $\mu$m, as a function of the interwell barrier $V_b$ and the tilt, $\Delta$,
between the two wells. The blue (red) surfaces describe the spin-up (spin-down) 
probability when a spin-down (spin-up) fermion is assumed to be at the fixed point. 
(a,b,c,d,e) $\Delta=0$ [see schematic in (II)] and $V_b=11.14$ kHz (lower barrier). 
(f,g,h,i) $\Delta=0$ [see schematic in (II)] and $V_b=24.30$ kHz (high barrier). 
(j,k,l,m) $\Delta=0.5 \hbar \omega_y$ [see schematic in (IV)] and $V_b=24.30$ kHz.  
(n,o,p,q) $\Delta=2.5 \hbar \omega_y$ [see schematic in (VI)] and $V_b=24.30$ kHz.
(VII) gives schematically the spin function for the ground state with $S=1/2$. 
In all cases, the confinement frequencies of the 1D traps are 
$\hbar \omega_x=6.6$ kHz and $\hbar \omega_y=1$ kHz. The SPDs and CPDs in (b,c,d,g,h,k,l,o,p) correspond to the 
$S=1/2$, $S_z=1/2$ CI ground state [brown curve in the associated spectra (a,f,j,n)] at the point 
(marked by a star) $-1/g=-0.1 (\sqrt{2}l_{0y} \hbar \omega_y )^{-1}$;
$g$ here is the 1D contact-interaction strength along the $y$ direction \cite{yann15}. 
The fixed point (see black arrows) in the SR-CPDs is placed at 
${\bf r}_0=(+1.3$ $\mu$m,$-1.1$ $\mu$m) in (c,h),
${\bf r}_0=(-1.3$ $\mu$m,$-1.1$ $\mu$m) in (d,g,l), and
${\bf r}_0=(-1.3$ $\mu$m,0) in (p).  (VIII) shows schematically the degeneracies (lifting of degeneracies) 
in the two uncoupled (coupled) Heisenberg rings ($t$-$J$ model) corresponding to the spectra in (a,f). 
(e,i,m,q) The von Neumann entanglement entropies, calculated from the single-particle density matrix
\cite{yann07,yann15}. 
Note the increased or constant entanglement with increasing repulsion, and the larger values for the symmetric DW 
congurations (e,i) compared to the nonsymmetric (tilted) ones (m,q).
The zero of energy in the spectra corresponds to the ground-state total energy of the corresponding non-interacting 
system, that is, to 11.14 $\hbar \omega_y$ in (a), 12.62 $\hbar \omega_y$ in (f), 13.11 $\hbar \omega_y$ 
in (j), and  13.62 $\hbar \omega_y$ in (n). The presence of the interwell barrier in (n) accounts for the difference 
from the single-well value of $(5 \hbar \omega_y + 3 \hbar \omega_x)/2=12.40 \hbar \omega_y$.
}
\label{specn3}
\end{figure}

\begin{figure}[t]
\centering\includegraphics[width=7.5cm]{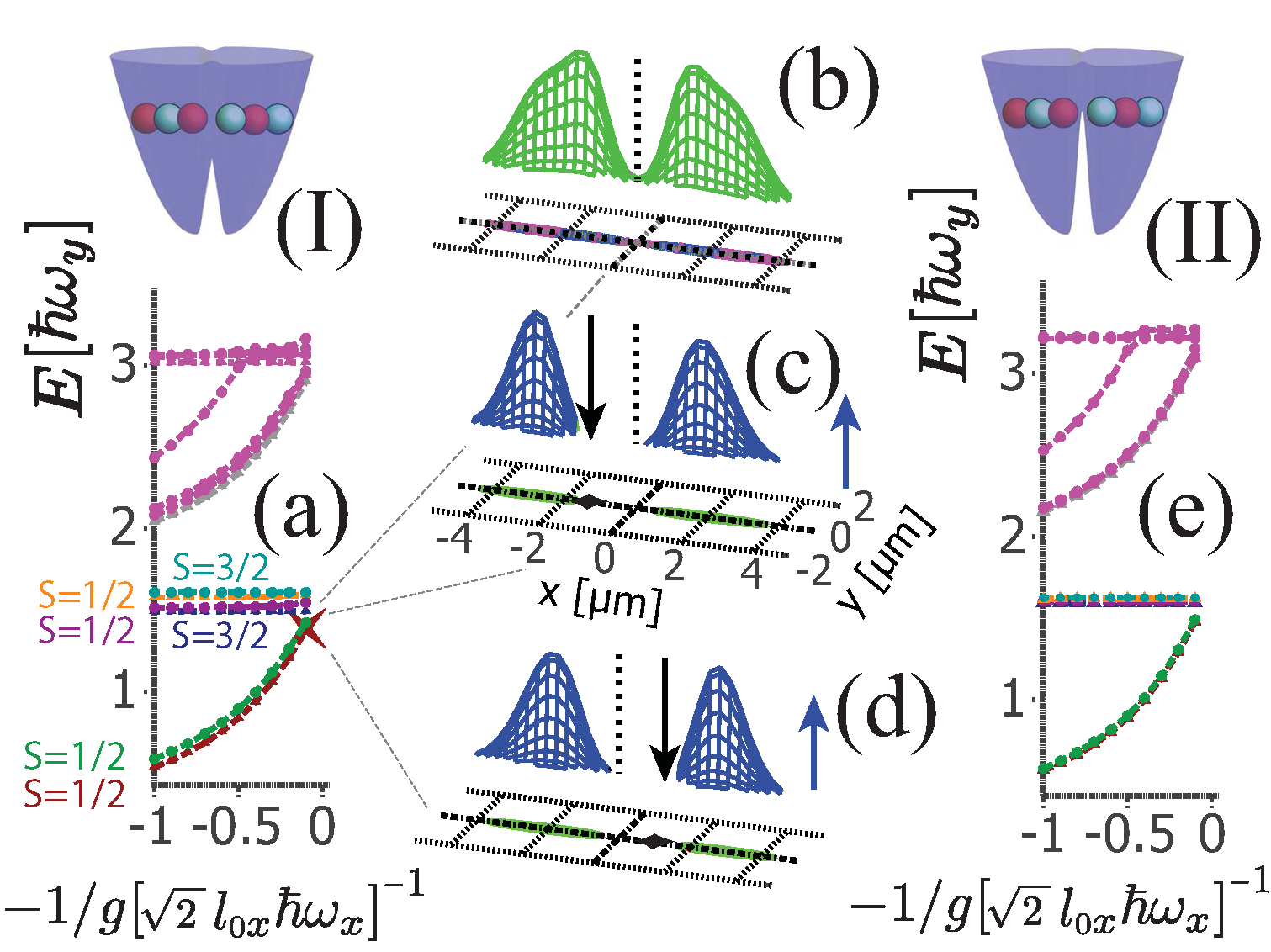}
\caption{Spectra, density, spin-resolved CPDs and entanglement entropies of 3 fermionic atoms in linear 
double-well confinements. (a) Energy versus $-1/g$ spectra, SPDs (green surfaces), (b) SPD (green surface), 
and (c,d) spin-resolved CPDs (blue surfaces) of $N=3$ strongly-repelling $^6$Li atoms in a symmetric $(\Delta=0)$ 
double-well confinement with a {\it linear\/} arrangement of the two 1D traps
at an interwell barrier $V_b=6.08$ kHz and $d=2.5$ $\mu$m [see schematic (I)].
The fermionic atoms are restricted to the lowest energy level in the $y$-direction and move within each well in the
$x$-direction ($\hbar \omega_y >> \hbar \omega_x)$, with 1D tunneling between the wells occuring along
the $x$-axis of the DWLI trap. (e) Corresponding spectrum for a very high barrier $V_b=11.14$ kHz at $d=2.5$ $\mu$m
[see schematic (II)], which displays a characteristic 2-4 degeneracy pattern.
In all instances $\hbar \omega_x=1$ kHz and $\hbar \omega_y=100$ kHz. The SPD in (b) and CPDs
in (c,d) correspond to the $S=1/2$, $S_z=1/2$ CI ground state [brown curve in the associated spectrum (a)] at the 
point (marked by a star) $-1/g=-0.1 (\sqrt{2}l_{0x} \hbar \omega_x )^{-1}$;
$g$ here is the 1D contact-interaction strength along the $x$ direction \cite{yann15}.
The CPD in (c,d) display the distribution of the two up spins when the fixed spin-down fermion (see the black arrow)
is placed at ${\bf r}_0=(-1.3$ $\mu$m, 0) and ${\bf r}_0=(+1.3$ $\mu$m, 0), respectively.
The different-color balls in (I) and (II) indicate the two resonating linear UCWMs.
The zero of energy in the spectra corresponds to the ground-state total energy of the corresponding
non-interacting system, that is, to 152.51 $\hbar \omega_x$ in (a) and 152.73 $\hbar \omega_x$ in (b).
}
\label{specs6}
\end{figure}

In all instances, i.e., for both the DWPA and DWLI cases, as well as the SW case, there are two states with total 
spin $S=0$, three states with $S=1$, and one state with $S=2$. These total-spin multiplicities, denoted here by
${\cal G}(N=4,S)$, arise from the group-symmetry properties of the spin eigenfunctions 
\cite{yann09,pauncz} of $N=4$ fermions with spin $1/2$; for the multiplicities ${\cal G}(N,S)$ of total-spin 
degeneracies for any $N$ fermions, see the branching diagram \cite{pauncz} in \ref{speig}. 
(The theory of spin-1/2 eigenfunctions is well known in quantum chemistry (see Ref.\ \cite{pauncz}) and has been
used \cite{yann09} previously in the field of quantum dots, ant it will not be repeated here. However, for
a brief outline and a description of the general spin eigenfunctions for $N=3$ and $N=4$, see \ref{speig}.
Importantly, in the cases studied in this paper [that is, for $N=4$ (Figs.\  \ref{specpa} and \ref{specli}), 
and for $N=3$ in a SW and in the DWPA trap (Fig.\ \ref{specn3}), as well as in a DWLI trap (Fig.\ \ref{specs6})] 
the AFM lowest spin-state is the ground state and the energy level spacings decrease with increasing interatomic 
repulsion.

The similar behavior of the sixfold-multiplet bands irrespective of the different geometries of the double-well 
traps, i.e., DWPA versus DWLI, indicates an underlying physical process independent of dimensionality (2D 
versus 1D). This underlying physics involves spatial localization of the $^6$Li atoms at 
extemporaneously created sites within each well and the ensuing formation of quantum UCWMs, as can be seen by
an inspection of corresponding single-particle densities (SPDs, green surfaces in Figs.\ \ref{specpa} and 
\ref{specli}) and spin-resolved conditional probability distributions (SR-CPDs, angle-resolved pair correlations, 
red surfaces in Figs.\ \ref{specpa} and \ref{specli}). 

The SPD is the expectation value of the one-body operator
\begin{equation}
\rho({\bf r}) = \langle \Phi^{{\rm CI}}_N 
\vert  \sum_{i=1}^N \delta({\bf r}-{\bf r}_i)
\vert \Phi^{{\rm CI}}_N \rangle,
\label{spden}
\end{equation}
where $\vert \Phi^{{\rm CI}}_N \rangle$ denotes the many-body (multi-determinantal) CI wave function.

We note that the SPD is the sum of the spin-up and spin-down single-particle densities, defined as
\begin{equation}
\rho_\sigma ({\bf r}) = \langle \Phi^{{\rm CI}}_N
\vert  \sum_{i=1}^N \delta({\bf r}-{\bf r}_i) \delta_{\sigma \sigma_i}
\vert \Phi^{{\rm CI}}_N \rangle,
\label{spspin}
\end{equation}
where $\sigma$ and $\sigma_i$ denote up or down spins.

In all cases the SPDs display four humps corresponding to the four localized fermions at the 
self-generated localization sites. The detailed arrangement of these sites varies in order to accomodate 
the geometry of the traps. For the DWPA case [Fig.\ \ref{specpa}(f)] with two fermions in the left well and the 
other two in the right well ($n_L=2$, $n_R=2$), a 2D rectangle is formed. For the DWLI (2,2) case 
[Fig.\ \ref{specli}(c)], including the limiting case of the single well [Fig.\ \ref{specpa}(b)], the four sites fall 
onto a straight line. Note the opening in the middle of the DWLI density [Fig.\ \ref{specli}(c)], in contrast to the
case of the single well in Fig.\ \ref{specpa}(b).  

Although several distinct spin structures can correspond to the same SPD of a UCWM, the spin eigenfunction 
associated with a specific CI wave function can be determined with the help of the many-body SR-CPDs, 
${\cal P}_{\sigma\sigma_0}$, which yield the conditional probability distribution of finding another fermion with up 
(or down) spin $\sigma$ at a position ${\bf r}$, given that a specific fermion with up (or down) spin $\sigma_0$ is 
fixed at ${\bf r_0}$. In detail, the spin-resolved two-point anisotropic correlation function is defined as
\begin{equation}
P_{\sigma\sigma_0}({\bf r}, {\bf r}_0)= 
\langle \Phi^{{\rm CI}}_N |
\sum_{i \neq j} \delta({\bf r} - {\bf r}_i) \delta({\bf r}_0 - {\bf r}_j)
\delta_{\sigma \sigma_i} \delta_{\sigma_0 \sigma_j}
|\Phi^{{\rm CI}}_N \rangle.
\label{tpcorr}
\end{equation}
Using a normalization constant 
\begin{equation}
{\cal N}(\sigma,\sigma_0,{\bf r}_0) = 
\int P_{\sigma\sigma_0}({\bf r}, {\bf r}_0) d{\bf r},
\label{norm}
\end{equation}
we further define a related spin-resolved conditional probability distribution (SR-CPD) as
\begin{equation}
{\cal P}_{\sigma\sigma_0}({\bf r}, {\bf r}_0) =
P_{\sigma\sigma_0}({\bf r}, {\bf r}_0)/{\cal N}(\sigma,\sigma_0,{\bf r}_0).
\label{cpd}
\end{equation}

In particular, by calculating
the ratios of the volumes under the CPD humps and equating them to the corresponding ratios 
of the squares of the angle-dependent coefficients of the general expressions for the spin eigenfunctions, one can 
determine the numerical values of the coefficients that map the spin eigenfunction to a specific 
SR-CPD (for details, see \ref{speig} and Ref.\ \cite{yann09}). 
As an example, the spin eigenfunction associated with the 4-fermion $S=0$, $S_z=0$ CI ground state at $-1/g=-0.1  
(\sqrt{2}l_{0y} \hbar \omega_y )^{-1}$ in the case of well-separated DWPA parallel wells [see star in Fig.\ 
\ref{specpa}(e); for the corresponding SR-CPD, see Fig.\ \ref{specpa}(g)] is given by 
[$\theta=-\pi/3$ in Eq.\ (\ref{x00})] 
\begin{equation}
{\cal X}_{00}^{(1)}=(-\alpha \alpha \beta \beta
+\alpha \beta \alpha \beta 
+\beta \alpha \beta \alpha
-\beta \beta \alpha \alpha)/2,
\label{x00_cpd1}
\end{equation}
where the $\alpha$'s ($\beta$'s) denote up (down) spin-$1/2$ fermions situated at the self-generated sites
(the maxima of the humps in the SPDs or CPDs); the methodology and detailed calculations used in determining the 
angle $\theta$ in the general spin eigenfunction in Eq.\ (\ref{x00}) are described in \ref{speig}.    

The equal in absolute value $|{\cal C}_i|=1/2$, $i=1,\ldots,4$ coefficients in front of the four primitives in 
Eq.\ (\ref{x00_cpd1}) agree with the probability of 0.5 [i.e., $0.5= 2 \times {\cal C}^2$] 
for the so-called ``antiferromagnetic'' component ($\alpha \alpha \beta \beta$ and $\beta \beta \alpha \alpha$)
found in Ref.\ \cite{zinn14} [see Fig.\ 1(d) therein] for the case of a two-parabola DWLI double well in the 
high-barrier regime. They also agree with the probability for the ``mixed'' component ($\alpha \beta \alpha \beta$ 
and $\beta \alpha \beta \alpha$) reported in the same paper. We note that in our treatment, 
we can vary the height of the barrier
independently from the separation of the wells, unlike the case in Ref.\ \cite{zinn14}. The use of the terms 
``antiferromagnetic'', ``mixed'', and ``ferromagnetic'' to characterize the spin primitives of the Ising basis
is borrowed here and in a paragraph below from Ref.\ \cite{zinn14} in order to facilitate the comparisons. This 
use is not repeated anywhere else in the paper; instead, as aforementioned, we employ the term ``antiferromagnetic'' 
to describe finite systems that have ground states with the lowest possible total spin.  

The mapping to the spin eigenfunction in Eq.\ (\ref{x00_cpd1}) reflects the fact that at the high-barrier (or
large-separation) regime the 4-fermion problem can be viewed as that of two pairs of strongly interacting
fermions within each well, each pair interacting weakly with the other one through the high barrier. 
In this case, as discussed below, the energetics of the 4-fermion system can be understood simply by adding
the singlet and triplet energy levels of the left and right fermionic pairs. However, the CI wave functions exhibit 
strong entanglement between the left- and right-well fermionic pairs in addition to the entanglement between the 
two fermions within each well. This across-the-barrier entanglement is not weakening as a result of a higher barrier,
and it is manifested in the mapping of the CI ground-state wave function onto the spin eigenfunction 
in Eq.\ (\ref{x00_cpd1}).

Furthermore, the discussion above applies also to the excited states. For example, the SPD and SR-CPD of the first 
excited state with $S=0$, $S_z=0$ in the DWPA trap of Fig.\ \ref{specpa} [having an energy ${\cal E}=2$ $\hbar 
\omega_y$ in Fig.\ \ref{specpa}(e)] is displayed in Figs.\ \ref{specpa}(h) and \ref{specpa}(i), repectively.
For this case, following an analysis as described above (and in \ref{speig}), we find an angle $\theta = \pi/6$,
which is associated with a spin function of the form 
\begin{equation}
 {\cal X}_{00}^{(2)} = \frac{1}{2\sqrt{3}}\; \alpha \alpha \beta \beta + 
                       \frac{1}{2\sqrt{3}} \alpha \beta \alpha \beta -
                       \frac{1}{\sqrt{3}} \alpha \beta \beta \alpha  +
                       ( \alpha \leftrightarrow \beta ).
\label{x002}
\end{equation}
We note that the spin eigenfunctions in Eqs.\ (\ref{x00_cpd1}) and (\ref{x002}) are orthogonal.

The two coefficients ${\cal C}_1={\cal C}_2=1/(2\sqrt{3})$ in front of the first two 
primitives in Eq.\ (\ref{x002}) agree with the probability of 0.166 (i.e., $0.166= 2 \times {{\cal C}_1}^2$) found
in Ref.\ \cite{zinn14} for the ``antiferromagnetic'' ($\alpha \alpha \beta \beta$ and $\beta \beta \alpha \alpha$), 
as well as for the ``mixed'' ($\alpha \beta \alpha \beta$ and $\beta \alpha \beta \alpha$) primitives in the case of 
a two-parabola DWLI double well at the high-barrier regime. The third coefficient  ${\cal C}_3=-1/\sqrt{3}$ for the 
``ferromagnetic'' primitive in Eq.\ (\ref{x002}) yields a probability of 0.666 ($0.666 = 2 \times {{\cal C}_3}^2$), 
again in agreement with Ref.\ \cite{zinn14}.

The spin eigenfunction associated with the 4-fermion $S=0$, $S_z=0$ CI ground state at 
$-1/g=-0.1 (\sqrt{2}l_{0x} \hbar \omega_x )^{-1}$ in the case of well-separated wells in the DWLI linear 
configuration [see star in Fig.\ \ref{specli}(b); for the corresponding SR-CPD, see Fig.\ \ref{specli}(d)] is 
given by the same spin eigenfunction as in Eq.\ (\ref{x00_cpd1}). This is due to the fact that the left and right 
pairs of fermions are isolated from each other in their respective wells. 

Returning to the case of four fermions in a single quasi-1D trap [Figs.\ \ref{specpa}(a,b,c)], 
the spin eigenfunction associated with the $S=0$, $S_z=0$ CI ground state at 
$-1/g=-0.1 (\sqrt{2}l_{0y} \hbar \omega_y )^{-1}$ [see star in Fig.\ \ref{specpa}(a); for the corresponding SR-CPD, 
see Fig.\ \ref{specpa}(c)] is found to have a different form from those in Eqs.\ (\ref{x00_cpd1}) and (\ref{x002}).
Specifically, the analysis of the SR-CPD described in detail in \ref{speig} yields an angle $\theta=-\pi/5.12$ in 
Eq.\ (\ref{x00}), which is associated with the following spin eigenfunction
\begin{equation}
 {\cal X}_{00}^{(3)} =  
     C_1 \alpha \alpha \beta \beta 
   + C_2 \alpha \beta \alpha \beta 
   + C_3\alpha \beta \beta \alpha 
   + ( \alpha \leftrightarrow \beta ),
\label{x00_cpd2}
\end{equation}
where $C_1=0.332411$, $C_2=-0.575017$, and $C_3=0.242606$.

In the next section, we utilize the trends uncovered by the CI solutions for the spectra and wave functions of the
many-body Hamiltonian, in order to develop a Heisenberg-model phenomenology. This development aims at providing
tools for analyzing quantum magnetism in double (and multi-well) ultracold-atom traps.

\subsection{Four fermionic ultracold atoms: The Heisenberg model}
\label{4fhh}

We have verified that the CI energy spectra presented in Figs.\ \ref{specpa} and \ref{specli}, as well as the 
SR-CPD-derived spin eigenfunctions [see, e.g., the functions in Eqs. (\ref{x00_cpd1}), (\ref{x002}) and 
(\ref{x00_cpd2})] are related to those of a 4-site Heisenberg Hamiltonian ${\cal H}_H$, 
with the four fermions being located at the 
humps of the SPDs and SR-CPDs, namely to [see, e.g., Eqs.\ (\ref{hh1}) and (\ref{hh2})] 
\begin{equation}
{\cal H}_H = \sum_{<ij>} J_{ij} {\bf S}_i {\bf \cdot S}_j - \sum_{<ij>} J_{ij}/4,
\label{heih} 
\end{equation}
where the symbol $<ij>$ denotes that the summation is restricted to the nearest-neighbor sites. The second term
is a scalar, leading simply to an overall energy shift; for a detailed description of 
${\cal H}_H$, see \ref{ab} and \ref{ac}. The DWPA case
is associated with a rectangular 2D Heisenberg ring [see schematic (III) in Fig.\ \ref{specpa}], while the DWLI and 
SW cases represent open linear spin chains [see schematic (III) in Fig.\ \ref{specli}]. Due to the $x$ and $y$ 
reflection symmetries, ${\cal H}_H$ has only two different exchange constants. In particular, in general,
for the rectangular Heisenberg ring in the DWPA case, the interwell exchange constants $J_{12}=J_{34}=r\neq 0$ and 
the intrawell ones $J_{23}=J_{14}=s \neq 0$. 
For the open 1D linear configuration of the DWLI and SW traps, $J_{12}=r \neq 0$, $J_{34}=0$, 
and $J_{23}=J_{14}=s \neq 0$. The energy eigenvalues ${\cal E}_i$ and eigenvectors 
${\cal V}_i$ of ${\cal H}_H$ [Eq.\ (\ref{heih})] are given in  \ref{ab} and \ref{ac}. They can 
reproduce all the trends in the energy spectra of the sixfold energy band, as well as the total-spin multiplicities 
${\cal G}(N=4,S)$ and spin eigenfunctions calculated via the CI method. In particular, in the limit of 
well-separated wells (i.e., for $r=0$), one gets ${\cal E}_2={\cal E}_4={\cal E}_6=0$, ${\cal E}_1={\cal E}_3=-s$, 
and ${\cal E}_5=-2s$, which coincides with the aforementioned 1-2-3 spin-group-theoretical degeneracy 
pattern and relative gaps within the sixfold lowest-energy CI band. 
Note further that the Heisenberg modeling reproduces the two different SR-CPD-derived spin eigenfunctions in 
Eqs.\ (\ref{x00_cpd1}) and (\ref{x002}), associated with the fully-separated-wells ($r=0$, for both the DWPA and 
DWLI cases); compare with the eigenvectors in Eqs.\ (\ref{v52}) and (\ref{v2_5}). 

It is notable that both the CI spectra [see Figs.\ \ref{specpa}(e) and \ref{specli}(b)] and the Heisenberg
energies for fully separated wells exhibit two energy gaps, one twice as large as the other (e.g., $-s$ and
$-2s$ in the Heisenberg model). This behavior can be understood from the spectrum of two unrelated single wells 
each containing a pair of two strongly interacting fermions. Indeed, the two lowest levels of two interacting fermions
consist of a singlet state with energy $E_s$ and a triplet state with energy $E_t$. The low-energy 
spectrum of the double well has then three levels, ${\cal E}_1=2E_t$, ${\cal E}_2=E_t+E_s$, and ${\cal E}_3=2E_s$, 
corresponding to whether both fermion pairs are in a triplet state, one pair is in a triplet with the other in a 
singlet state, or both pairs are in a singlet state; this results in the two energy gaps 
$\Delta {\cal E}_{12}=E_t-E_s$ and $\Delta {\cal E}_{13}=2(E_t-E_s)=2 \Delta {\cal E}_{12}$.

The topology of the spin chain in Fig.\ \ref{specpa}(III) (DWPA) is indeed a closed ring, whereas the one in 
Fig.\ \ref{specli}(III)  (DWLI) is that of an open ring. The corresponding Heisenberg Hamiltonians are given in 
Eqs.\ (\ref{hh1matgen}) and (\ref{hh2mat}), respectively; note that they have different matrix elements. 
The similarities between these two cases arise from the fact that the spin eigenfunctions onto which the CI 
wavefunctions map (as we show in both the DWPA and DWLI cases) have the same group structure, differing only in the 
coefficients of their components [see, e.g., Eq.\ (\ref{x00}) in  \ref{speig}]; the multiplicity of the four fermions 
spin eigenfunctions onto which the CI spectrum maps (in both the DWPA and DWLI cases) is six (for all arrangements of 
4 fermions, see \ref{speig} and Fig.\ \ref{brdi}). 

For the single-well case, all six CI energies have distinct values; see the spectrum in Fig.\ \ref{specpa}(a). By
using the open-Heisenberg-chain eigenvalues ${\cal E}_i$, $i=1,\ldots,6$ in Eqs.\ (\ref{e2_1}-\ref{e2_6}) and fitting 
the ratios $({\cal E}_4-{\cal E}_i)/({\cal E}_4-{\cal E}_j)$ to the CI spectrum, we can determine the parameter
$f=r/s$ that describes the single well. For example, using the fully polarized, $E_{fp}={\cal E}_4=0$, the 
ground-state, $E_{gs}={\cal E}_5$, and the 1st-excited, $E_{1st}={\cal E}_3$, energies, we obtain the ratio
\begin{equation}
\frac{E_{1st}-E_{fp}}{E_{gs}-E_{fp}} = \frac{\sqrt{f^2+1}+f+1}{f+\sqrt{(f-2)f+4}+2},
\label{rsfit}
\end{equation}
which is independent of $s$ and allows for the determination of $f$. Fitting to the CI spectrum, we get $f \sim 1.35$.
This value agrees with that resulting from the nearest-neighbor exchange constants of
harmonically trapped particles listed in Table I of Ref.\ \cite{deur14}. Another study \cite{zinn15} gave a value
of $\approx 1.4$ for this ratio.

With the value of $r/s=1.35$, the open linear Heisenberg chain yields ${\cal E}_1=-s$ ($S=1$), 
${\cal E}_2=-0.334985 s$ ($S=1$), ${\cal E}_3=-2.01501 s$ ($S=1$), ${\cal E}_4=0$
($S=2$),  ${\cal E}_5=-2.55853 s$ ($S=0$), and ${\cal E}_6=-0.79147 s$ ($S=0$), 
i.e., six distinct values, in agreement with the CI spectrum in Fig.\ \ref{specpa}(a).  
The corresponding angle in Eq.\ (\ref{x00}) is $\theta = -\pi/4.58$. This value is slightly different from the value 
of $-\pi/5.12$ (corresponding to an $r/s \approx 1.62$) that was determined above in Sec.\ \ref{4fci} from an analysis
of the CI CPD in Fig.\ \ref{specpa}(a). This slight discrepancy is due to the elimination of the space degrees of 
freedom when considering the mapping of the CI wave function onto the spin eigenfunctions. Naturally, the spin
eigenfunctions have constant coefficients in front of the Ising-expansion primitives and by themselves are unable to 
reflect the influence of the extent of space distribution of the localized fermions. Indeed the localization of
the four fermions is sharper in a double well with a high barrier compared to that in a single well; compare
the SPD's in Figs.\ \ref{specpa}(b) (4 fermions inside the same well) and \ref{specpa}(f) (2 fermions in each
well). The CI SPDs and SR-CPDs incorporate in their definition the space degrees of freedom and they account for the
actual extent of partial or full particle localization (which varies with $g$). A detailed investigation of this 
matter is beyond the scope of this paper, but it will be examined in a future publication \cite{yannf}.  


\section{Three fermionic ultracold atoms in a double-well trap}
\label{3ferm}

\subsection{Three fermionic ultracold atoms: CI results and the Heisenberg model for tilted double wells}
\label{3fci}

CI results for $N=3$ ultracold $^6$Li atoms in a DWPA trap are displayed in Fig.\ \ref{specn3} for both 
symmetric [zero tilt, $\Delta=0$, see schematic in Fig.\ \ref{specn3}(II) and spectra in Fig.\ \ref{specn3}(a) and 
Fig.\ \ref{specn3}(f)] and asymmetric wells with a moderate tilt $\Delta=0.5 \hbar \omega_y$ [see 
Fig.\ \ref{specn3}(IV) and Fig.\ \ref{specn3}(j)] and a strong tilt $\Delta=2.5 \hbar \omega_y$ [see
Fig.\ \ref{specn3}(VI) and Fig.\ \ref{specn3}(n)]. 

The cases of asymmetric wells are amenable to straightforward interpretations based on pure Heisenberg models.
The moderate tilt [Fig.\ \ref{specn3}(IV), $\Delta=0.5 \hbar \omega_y$] generates a ground state with a (2,1) 
distribution of the atoms (two in the left well and one in the right, tilted upward, one),
which are localized in the shape of a isosceles triangular UCWM 
[see the SPD in Fig.\ \ref{specn3}(k)]. 
The corresponding CI energy spectrum [Fig.\ \ref{specn3}(j)] exhibits a three-fold lowest-energy band with
a characteristic 1-2 degeneracy pattern, converging to the same energy for $-1/g \rightarrow 0$. The total-spin 
multiplicities in this band are ${\cal G}(N=3,S=1/2)=2$ and ${\cal G}(N=3,S=3/2)=1$, in agreement 
with the branching diagram for three fermions (see \ref{speig}). 
This CI energy spectrum and the correponding SR-CPDs 
[see, e.g., the SR-CPD in Fig.\ \ref{specn3}(l)] are reproduced by a 3-site Heisenberg-ring Hamiltonian
\begin{equation}
{\cal H}_H^{\rm trg}=J_{12}  {\bf S}_1 {\bf \cdot S}_2 + J_{13} ({\bf S}_1 {\bf \cdot S}_3 +
{\bf S}_2 {\bf \cdot S}_3) - J_{12}/4 - J_{13}/2,
\label{heihtrg}   
\end{equation}
with $J_{12}=s$ and $J_{13}=J_{23}=r$; for the numbering of the three sites, see the schematics in Fig.\ 
\ref{specn3}(V) and in \ref{heis3}. For $r=0$ (case of a high barrier $V_b$), the
eigenenergies of ${\cal H}_H^{\rm trg}$ are ${\cal E}_1\;(S=3/2)={\cal E}_2 \; (S=1/2)=0$ and 
${\cal E}_3 \; (S=1/2)=-s$, reproducing the above-mentioned 1-2 CI degeneracy pattern. The CI-calulated CPDs are 
also in full agreement with the eigenvectors of the ${\cal H}_H^{\rm trg}$ Hamiltonian. For example, the CI 
ground-state SR-CPD ${\cal P}_{\downarrow\uparrow}$ in Fig.\ \ref{specn3}(l) 
[at the point $-1/g=-0.1 (\sqrt{2}l_{0y} \hbar \omega_y )^{-1}$] is found to map 
onto the 3-fermion general spin eigenfunction 
[see Eq.\ (\ref{x1212})] 
for $\theta=0$, i.e., that is to the function
\begin{equation}
{\cal X}_{1/2,1/2}^{(1)}= (\alpha \beta \alpha - \beta \alpha \alpha)/\sqrt{2}.
\label{x1212gs1}
\end{equation}
This CI-derived spin function is schematically portrayed in Fig.\ \ref{specn3}(V) and agrees with the Heisenberg
eigenvector in Eq.\ (\ref{v33_3}). 

\begin{figure}[t]
\centering\includegraphics[width=13cm]{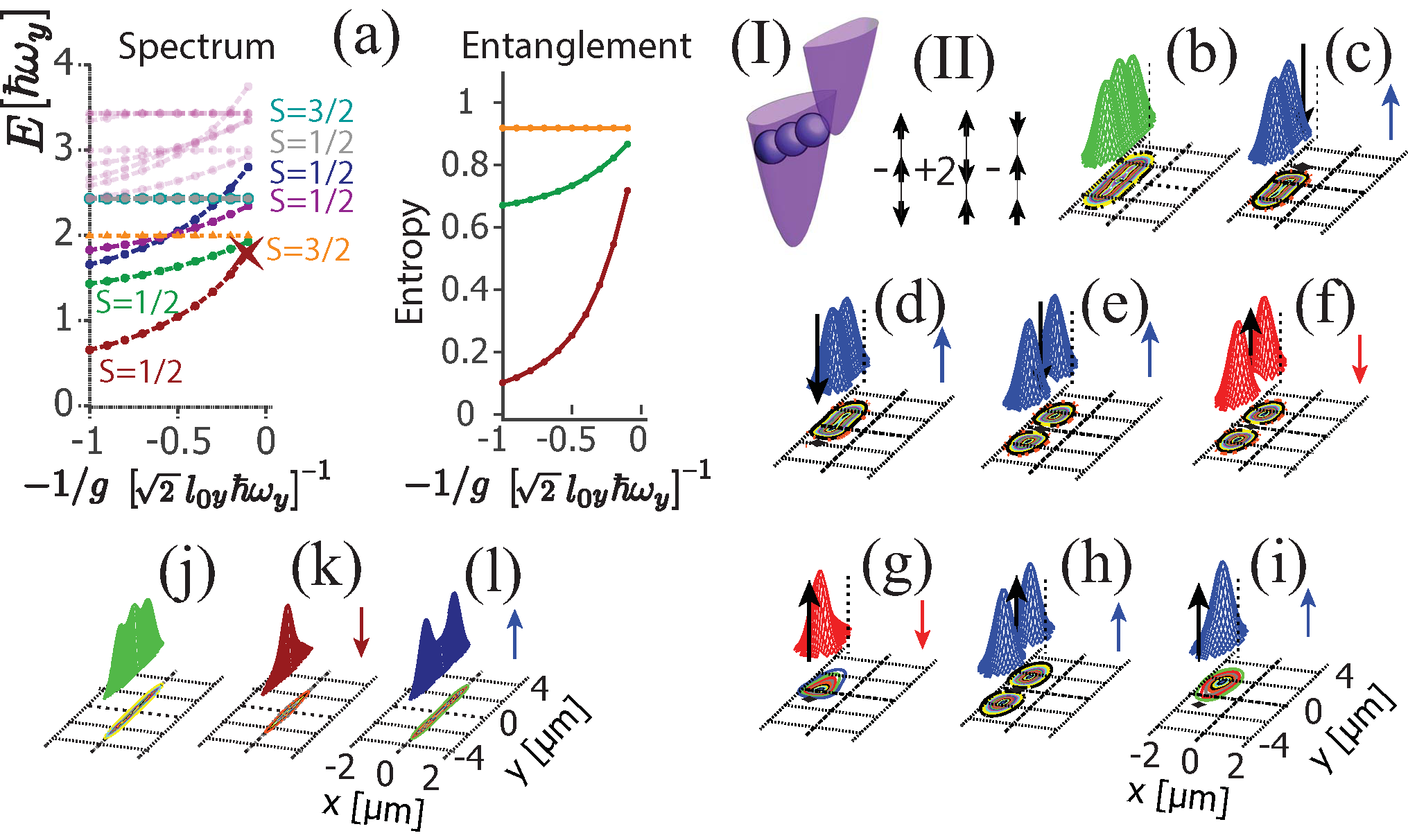}
\caption{(a-i) Spectra, entanglement, total densities, and spin-resolved CPDs of $N=3$ 
strongly-repelling $^6$Li atoms in a 1D single-well harmonic confinement 
[as a limiting case of a DWPA for strong tilt $\Delta=2.5$ kHz and 
strong interwell barrier $V_b=24.30$ kHz; see schematic in (I)]). (a) Energy spectrum and von Neumann entropy 
versus $-1/g$. (b) SPD (green surface). (c-i) spin-resolved CPDs.
The blue (red) surfaces describe the spin-up (spin-down) probability when a spin-down (spin-up) fermion is assumed 
to be at the fixed point (denoted by a black vector). The confinement frequencies of the 1D traps are
$\hbar \omega_x=6.6$ kHz and $\hbar \omega_y=1$ kHz. The SPD (b) and SR-CPDs in (b-i) correspond to the
$S=1/2$, $S_z=1/2$ CI ground state [brown curve in the associated spectrum (a)] at the point
(marked by a star) $-1/g=-0.1 (\sqrt{2} l_{0y} \hbar \omega_y )^{-1}$;
$g$ here is the 1D contact-interaction strength along the $y$ direction \cite{yann15}.
The fixed point (see black arrows) in the SR-CPDs is placed at
${\bf r}_0=(-1.3$ $\mu$m,$1.9$ $\mu$m) in (c), ${\bf r}_0=(-1.3$ $\mu$m,$-1.9$ $\mu$m) in (d),
${\bf r}_0=(-1.3$ $\mu$m,$0$) in (e), ${\bf r}_0=(-1.3$ $\mu$m,$0$ ) in (f),
${\bf r}_0=(-1.3$ $\mu$m,$-1.1$ $\mu$m) in (g), ${\bf r}_0=(-1.3$ $\mu$m,$0$) in (h),
and ${\bf r}_0=(-1.3$ $\mu$m,$-1.1$ $\mu$m) in (i).
The von Neumann entanglement entropies in (a) are calculated from the single-particle density matrix
\cite{yann07,yann15}. This figure complements Fig.\ \ref{specn3} in that it displays examples of 
all possible SR-CPDs for the ground state of $N=3$ cold fermions in a single 1D well. It is 
straightforward to check that the SR-CPDs agree with a mapping of the CI ground state onto
the spin eigenfunction of the schematic (II). (j-l) Ground-state results for $N=3$ $^6$Li atoms in a strictly-1D 
single trap with $\hbar \omega_x=100$ kHz and $\hbar \omega_y=1$ kHz. (j) SPD (green surface). (k) spin-down
density (red surface). (l) spin-up density (blue surface). 
The zero of energy in the spectrum corresponds to the ground-state total energy of the corresponding
non-interacting system, that is, to 13.62 $\hbar \omega_y$ in (a).
}
\label{specs4}
\end{figure}

A larger tilt $\Delta=2.5 \hbar \omega_y$ generates a (3,0) CI ground state, associated with a
linear UCWM [see the SPD in Fig.\ \ref{specn3}(o)]. 
The CI energy spectrum [Fig.\ \ref{specn3}(n)] and the correponding CPDs [see, e.g., Fig.\ \ref{specn3}(p)]
are related to a 3-site open-linear-chain Heisenberg Hamiltonian, obtained from Eq.\ (\ref{heihtrg}) by setting 
$J_{12}=s=0$. This Hamiltonian has three different eigenenergies 
${\cal E}_1\; (S=3/2)=0$, ${\cal E}_2\; (S=1/2)=-3r/2$, and ${\cal E}_3\; (S=1/2)=-r/2$, in agreement with
the threefold CI band.
The ground-state CI-derived spin function is schematically portrayed in Fig.\ \ref{specn3}(VII) [$\theta=\pi/2$ in 
Eq.\ (\ref{x1212})] and agrees with the Heisenberg eigenvector in Eq.\ (\ref{v33_2}). 

Fig.\ \ref{specs4} complements Fig.\ \ref{specn3} in that it displays examples of all possible SR-CPDs for the ground 
state of $N=3$ cold fermions in a single 1D well. It is straightforward to check in detail that the all SR-CPDs agree 
with a mapping of the CI ground state onto the spin eigenfunction of the schematic in Fig.\ \ref{specs4}(I), 
i.e., with the Heisenberg vector ${\cal V}_2$ in Eq.\ (\ref{v33_2}). We note that, while the spin spatial distribution
is analyzed here with the use of the SR-CPDs [see Eq.\ (\ref{tpcorr})], it is also reflected in the spatial 
spin-densities [see Eq.\ (\ref{spspin})] shown in Figs.\ \ref{specs4}(k-l); the latter agree with those displayed in 
Fig.\ 6 of Ref.\ \cite{deur14}. Note that the sum of the up- and down-spin densities in Figs.\ \ref{specs4}(k-l) 
agrees with the total SPD in Fig.\ \ref{specs4}(j). 

A qualitatively different behavior, bringing extra intricacies and opening igress to novel complex physical systems,
is exhibited by the symmetric DWPA cases ($\Delta=0$) for $N=3$ shown in Fig.\ \ref{specn3}. Indeed, the CI energy 
spectra in Fig.\ \ref{specn3}(a) and Fig.\ \ref{specn3}(f) show a sixfold lowest-energy band, comprising four 
$S=1/2$ states, and two $S=3/2$ states, i.e., twice as many as in the case of tilted wells [Figs.\
\ref{specn3}(j) and \ref{specn3}(n)]. In particular, for the higher barrier [Fig.\ \ref{specn3}(f)] a characteristic
2-4 degeneracy appears, which is a doubling of the 1-2 degeneracy pattern in Fig.\ \ref{specn3}(j).
This doubling of the number of energies is due to the conservation of parity, which requires consideration of a 
second triangle (246), which is the mirror of the original (135) one; see the schematic in Fig.\ \ref{specn3}(I) and 
in Fig.\ \ref{2trg}(a). 
In each of these mirror reflected  configurations, two atoms localize in one well and one atom localizes in the other
well; see the two sets of different colored spheres in Fig.\ \ref{specn3}(I). The formation of these triangular 
atomic configurations is reflected in the SR-CPDs shown on Fig.\ \ref{specn3}(c,d) for the lower-barrier symmetric 
DW case and Fig.\ \ref{specn3}(g,h) for the higher-barrier case. One may view this situation as having six available 
sites altogether (three in each well), with the 3 fermionic atoms localizing in either of the aforementioned 
triangular configurations, (135) and (246) [see Fig.\ \ref{2trg}(a)], with 2 atoms in one well and 1 atom in the 
other; in each case we may term the unoccupied (empty) sites as ``holes''. This mapping leads to the picture of a 
3-atom UCWM that resonates between the two interlocking triangles.

We mention that the resonating behavior and the symmetrization of the many-body wave function in 
two-center/three-electron bonded systems is well known \cite{hibe94,bick98,berr16} in theoretical chemistry and in 
particular in the valence-bond treatment of the three-electron bond which controls the formation of molecules like 
He$^+_2$ and F$_2^-$. Furthermore we mention that the symmetry properties of the strictly-1D few-fermion
problem with contact interactions have been also investigated in Refs.\ \cite{hars12,hars16} 

\subsection{Three fermionic ultracold atoms: The $t$-$J$ model for symmetric double wells}
\label{3ftj}

\begin{figure}[t]
\centering\includegraphics[width=5cm]{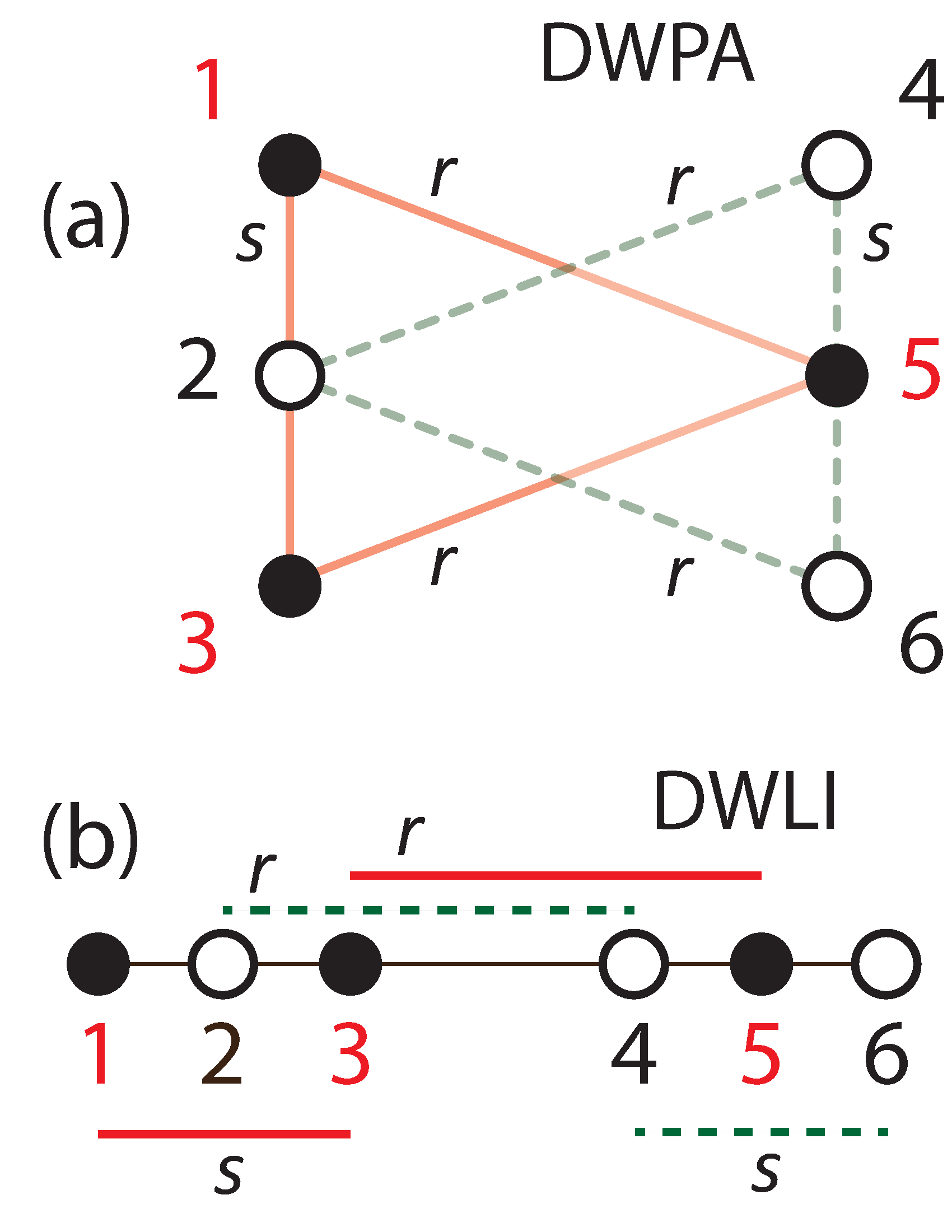}
\caption{Schematics of the six-site numbering conventions in the $t-J$ Hamiltonian used 
for 3 fermions in a symmetric double well (zero tilt). 
(a) DWPA arrangement and the associated two-interlocking-triangles geometry. (b) DWLI arrangement. 
}
\label{2trg}
\end{figure}

To model the exact-diagonalization results shown above, one must go beyond the aforementioned simple Heisenberg 
Hamiltonian model [see Eq.\ (\ref{heihtrg}) and \ref{heis3}]. 
Indeed, we find that a generalization of the  so called 
$t$-$J$ model allows us to capture all the salient characteristics uncovered by the  CI  calculations.  The $t$-$J$
model \cite{auerbookk,dago944} modifies (away from the half filling) the antiferromagnetic Heisenberg Hamiltonian 
associated with the Mott insulator at half-filling (one electron per crystal site); it has attracted much 
attention, because it has been proposed for explaining the high-T$_c$ superconductivity arising in the case of 
underdoped insulators (away from the half filling when holes are present). A finite $t$-$J$-type Hamiltonian may be 
expressed as 
\begin{equation}
{\cal H}_{tJ} = {\cal H}_H^{\rm trg}(135)(\{J\}) + {\cal H}_H^{\rm trg}(246)(\{J\}) + {\cal H}_c(\{t\}),
\label{htj}
\end{equation}   
where $H_c$ is the coupling between the two simple Heisenberg rings defined over the sites (135) and (246); see 
the two $3 \times 3$ blocks on the diagonal (upper left and lower right) in Eq.\ (\ref{htjm}).  
$H_c$, represented by the two off-diagonal blocks in Eq.\ (\ref{htjm}), is defined by the matrix
$t=(\alpha 0 \alpha 0 \beta 0 | {\cal H}_c | 0 \alpha 0 \alpha 0 \beta )=
(\alpha 0 \alpha 0 \beta 0 | {\cal H}_c | 0 \alpha 0 \beta 0 \alpha )$, and
$t_2=(\alpha 0 \alpha 0 \beta 0 | {\cal H}_c | 0 \beta 0 \alpha 0 \alpha )$, where the ``0'' indicates an 
empty site; e.g., $\alpha 0 \alpha 0 \beta 0$ corresponds to a state where sites 1,3, and 5 are occupied 
and 2,4, and 6 are empty (for site designation see Fig.\ \ref{specn3}(I) and Fig.\ \ref{2trg}). 

The Hamiltonian ${\cal H}_{tJ}$ is equivalent to a six by six matrix,\\
~~~~~~\\
{\scriptsize
\begin{equation}
\fl {\cal H}_{tJ}^\Delta =  
\left(
\begin{array}{ccc|ccc}
-J_{15}-\frac{\Delta}{2} & \frac{J_{15}}{2} & \frac{J_{15}}{2}   & t_2         & t                 & t         \\
\frac{J_{15}}{2}   & -\frac{J_{13}}{2}-\frac{J_{15}}{2}-\frac{\Delta}{2}   & \frac{J_{13}}{2} & t  & t  & t_2   \\
\frac{J_{15}}{2}  & \frac{J_{13}}{2}  & -\frac{J_{13}}{2}-\frac{J_{15}}{2}-\frac{\Delta}{2} & t  & t_2  & t \\ \hline
t_2   & t   & t   & -J_{24}+\frac{\Delta}{2} & \frac{J_{24}}{2}  & \frac{J_{24}}{2}     \\
t  & t  & t_2  & \frac{J_{24}}{2}  & -\frac{J_{46}}{2}-\frac{J_{24}}{2}+\frac{\Delta}{2} & \frac{J_{46}}{2}      \\
t   & t_2  & t  & \frac{J_{24}}{2}  & \frac{J_{46}}{2}  & -\frac{J_{46}}{2}-\frac{J_{24}}{2}+\frac{\Delta}{2}  \\
\end{array}
\right),
\label{htjm}
\end{equation}
}
~~~~~~\\
\noindent
where the upper index $\Delta$ denotes explicitly the dependence on the tilt. When $|\Delta| >> |t|$ and
$|\Delta| >> |t_2|$, one recovers the isolated-triangle Hamiltonian, ${\cal H}_H^{\rm trg}$, 
in Eq.\ (\ref{heihtrg}). Below we will focus on the case of symmetric double wells, i.e., we will set 
$\Delta=0$, $J_{15}=J_{24}=r$, and $J_{13}=J_{46}=s$. 

For $r=0$ and $t_2=t=0$ [case of the very high interwell barrier, $V_b=24.30$ kHz,  in Fig.\ \ref{specn3}(f)], 
${\cal H}_{tJ}^{\Delta=0}$ reproduces ($\times 2$) the characteristic CI 1-2 degeneracy pattern found earlier using 
the simple 3-site Heisenberg model [compare Fig.\ \ref{specn3}(f) and Fig.\ \ref{specn3}(j)]; see the six
eigenvalues ${\cal E}_i$, $i=1,\ldots,6$ in Eqs.\ (\ref{e4_1})-(\ref{e4_6}). 

For lower values of the interwell barrier [$V_b=11.14$ kHz, Fig.\ \ref{specn3}(a)], the 2-4 [(1-2) $\times 2$] 
doubling degeneracy is lifted, with two lowest $S=1/2$ curves and four higher 
in energy (and parallel) curves (two with $S=1/2$ and two with $S=3/2$) forming distinct subbands; then one 
distinguishes all 6 lines as separate lines [see the spectrum in Fig.\ \ref{specn3}(a) and also in
Fig.\ \ref{specs6}(a)]. 
It is remarkable that the nontrivial spectrum in Fig.\ \ref{specn3}(a) can be reproduced by setting 
$t_2 \sim - 4t/10 > -1/2 $, with $t<0$ and $s > |t|$. Then one has for the energy gap 
between the two lowest states, $\Delta {\cal E}_{12}={\cal E}_1-{\cal E}_2=14t/5$; these energies are centered around
$-s$. The remaining energies group together forming a fourfold band, centered around zero. The energy gap 
between the two outer (both $S=3/2$) members of the fourfold band is 
$\Delta {\cal E}_{56}={\cal E}_5-{\cal E}_6=16t/5$, i.e., similar to the $\Delta {\cal E}_{12}$ gap, 
in agreement again with the pattern in Fig.\ \ref{specn3}(a). Furthermore, 
the gap between the two higher energies in the fourfold band, as well as that between the two lower energies of
this band, is $\Delta {\cal E}_{35}=\Delta {\cal E}_{46}={\cal E}_4-{\cal E}_6=t/5$, which is much smaller 
than the width, $\Delta {\cal E}_{56}$, of the same band, again in agreement with the pattern in 
Fig.\ \ref{specn3}(a). Note that for $t_2=-t/2$, $\Delta {\cal E}_{35}=\Delta {\cal E}_{46}=0$ 
and a degeneracy pattern 1-1-2-2 develops in disagreement with the CI spectrum. 
Also, when $t_2=0$, the width $\Delta {\cal E}_{56}$ $(=4t)$ of the fourfold band is twice as large as the energy 
gap, $\Delta {\cal E}_{12}$ $(=2t)$, between the two lowest states, again in disageement with the CI spectrum in 
Fig.\ \ref{specn3}(a).

Similar trends pertaining to the doubling of the spectrum (from three to six states) in conjunction with the
emergence of two resonating UCWMs apply also in the case of $N=3$ ultracold fermions in a symmetric ($\Delta=0$) 
DWLI (linear arrangement) trap, as is illustrated in Fig.\ \ref{specs6}. These results can be interpreted again 
through the use of a corresponding $t$-$J$ model with a similar parametrization.

\section{Quantifying entanglement using a CI-based von Neumann entropy}
\label{enta}

The entanglement entropy $S_{vN}$ for three $^6$Li atoms in a DWPA trap in the configurations, whose spectra are 
shown in Fig.\ \ref{specn3}(a,f,j,n), are displayed in Fig.\ \ref{specn3}(e,i,m,q), respectively.

For the CI many-body wave functions, we adopt as a measure of entanglement the von Neumann entropy
\cite{yann07,yann15}, 
\begin{equation}
S_{vN}= - \Tr (\rho \log_2 \rho) +C,
\label{svn}
\end{equation}
where $\rho$ is the single-particle density matrix and  $C=-\log_2⁡N$, yielding $S_{vN}=0$ for an uncorrelated 
single-determinant state. 

The single-particle density matrix $\rho$ is given by
\begin{equation}
\rho_{\nu\mu} = \frac{\langle \Phi^{{\rm CI}}_N | a^\dagger_\mu a_\nu |
\Phi^{{\rm CI}}_N \rangle}
{\sum_\mu \langle \Phi^{{\rm CI}}_N | a^\dagger_\mu a_\mu |
\Phi^{{\rm CI}}_N \rangle}, 
\label{densmat}
\end{equation}
and it is normalized to unity, i.e., $\Tr \rho=1$. The Greek indices $\mu$ (or $\nu$) count the spin orbitals
(of dimension $2K$) 
\begin{equation}
\chi_j (x,y) = \varphi_j (x,y) \alpha, \mbox{~~~if~~~} 1 \leq j \leq K,
\label{chi1}
\end{equation}
and 
\begin{equation}
\chi_j (x,y) = \varphi_{j-K} (x,y) \beta, \mbox{~~~if~~~} K+1 \leq j \leq 2K, 
\label{chi2}
\end{equation}
where $\alpha (\beta)$ denote up (down) spins.

Since the allowed maximum value for $S_{vN}$ in our CI calculations is $\log_2(2K)-\log_2(3)=5.70$ (we use a typical 
basis of $K=78$ single-particle space orbitals), it is notable that the calculated values in Fig.\ \ref{specn3} remain
smaller than $\sim 1$, and in particular in the regime of strong correlations, i.e., for $-1/g \rightarrow -0$. 
This reflects formation of a Wigner molecule. Additionally, we find increased or constant entanglement with 
increasing repulsion, and larger values for the symmetric DW configurations [Fig.\ \ref{specn3}(e,i)] compared to the 
nonsymmetric (tilted) ones [Fig.\  \ref{specn3}(m,q)]. For $S_{vN}$ entropies for three $^6$Li atoms in a DWLI trap,
see Fig.\ \ref{specs4}(a).

\section{Summary and Outlook}
In this paper, we have presented timely advances in the growing field of few-body ultracold atoms with the aim of 
enhancing understanding of experimental endeavors and lodging new directions of research in this area. We progressed 
in two main courses: (i) uncovering universal non-itinerant and fermionization-like aspects of the physics of 
ultracold few fermions trapped in double-well confinements, with various 1D and 2D trapping geometries, as a conduit 
for emulating quantum magnetism and related phenomena beyond the strictly 1D single-well (SW) case, and (ii) making 
headways in the development and implementation of benchmark numerical simulations (exact diagonalization of the full 
microscopic Hamiltonian with configuration interaction, CI, techniques) as tools for modeling theoretical and 
experimental results with effective spin-Hamiltonians (Heisenberg and $t$-$J$ models). Our calculations for $N = 3$ 
and $N = 4$ ultracold fermionic $^6$Li atoms in SW and double well (DW) traps with linear (DWLI) or parallel (DWPA) 
geometries, reveal formation of antiferromagnetic ordering for the lowest-energy bands over the entire range of 
interparticle contact repulsion studied here. 

For $N = 4$ ultracold atoms in a symmetric DWPA trap with very strong interatomic repulsion, we find 
(via miscroscopic, CI, calculations) formation of a two-dimensional ultracold Wigner molecule (UCWM) of 
non-itinerant character. For the symmetric parallel DW trap the formation of the 2D UCWM leads to mapping of the 
interacting 4-atom trapped system onto a 2D rectangular Heisenberg ring cluster, whereas for a symmetric DWLI trap 
(as well as for a SW trap) we find a four-atom linear (1D) UCWM in juxtaposition with mapping onto a linear 
Heisenberg spin-chain. These mappings enable employment of the corresponding Heisenberg model Hamiltonian, whose 
solutions reproduce well the results of the microscopic, numerically-exact, calculations. 

For $N = 3$ ultracold atoms in DWLI or DWPA traps with a finite tilt (detuning) between the two wells, the 
numerically calculated (CI) spectrum for strong interatomic repulsion is described well with the use of the 
aforementioned Heisenberg Hamiltonian. As noted already in the Introduction, the high measure of 
entanglement predicted for the set of lowest energy states of the three strongly repelling fermionic atoms,  
together with the controllable tilt between the two wells, motivate consideration of this double well system as a 
cold-atom quantum computing qubit. 

In contrast to the asymmetric DW case, description of the $N=3$ ultracold-atom CI spectra for symmetric (vanishing 
tilt) DWLI or DWPA traps, that manifest doubling of the number of states in the lowest band, as well as modeling the 
corresponding SR-CPDs, are not attainable with the simple Heisenberg model, requiring instead the more intricate
$t$-$J$-type model \cite{auerbookk,fazebookk,dago944}, consisting of two coupled resonating triangular 2D UCWM 
Heisenberg clusters. 
The emergence of the $t$-$J$ model for the description of quantum magnetism (in particular AFM ordering) in a 
trapped few-body ultracold atom system, strongly suggests its future role as a useful laboratory for exploration of 
the elementary building blocks of high-T$_c$ superconducting behavior \cite{dago944,norm11,hule15}.

\ack
We acknowledge financial support from the Air Force Office of Scientific Research under Award 
No. FA9550-15-1-0519. Calculations were carried out at the GATECH Center for Computational Materials Science.\\

\setcounter{section}{1}
\appendix

\section{Spin eigenfunctions for 4 and 3 fermions. Comparison with CI CPDs}
\label{speig}

\begin{figure}[b]
\centering\includegraphics[width=5.5cm]{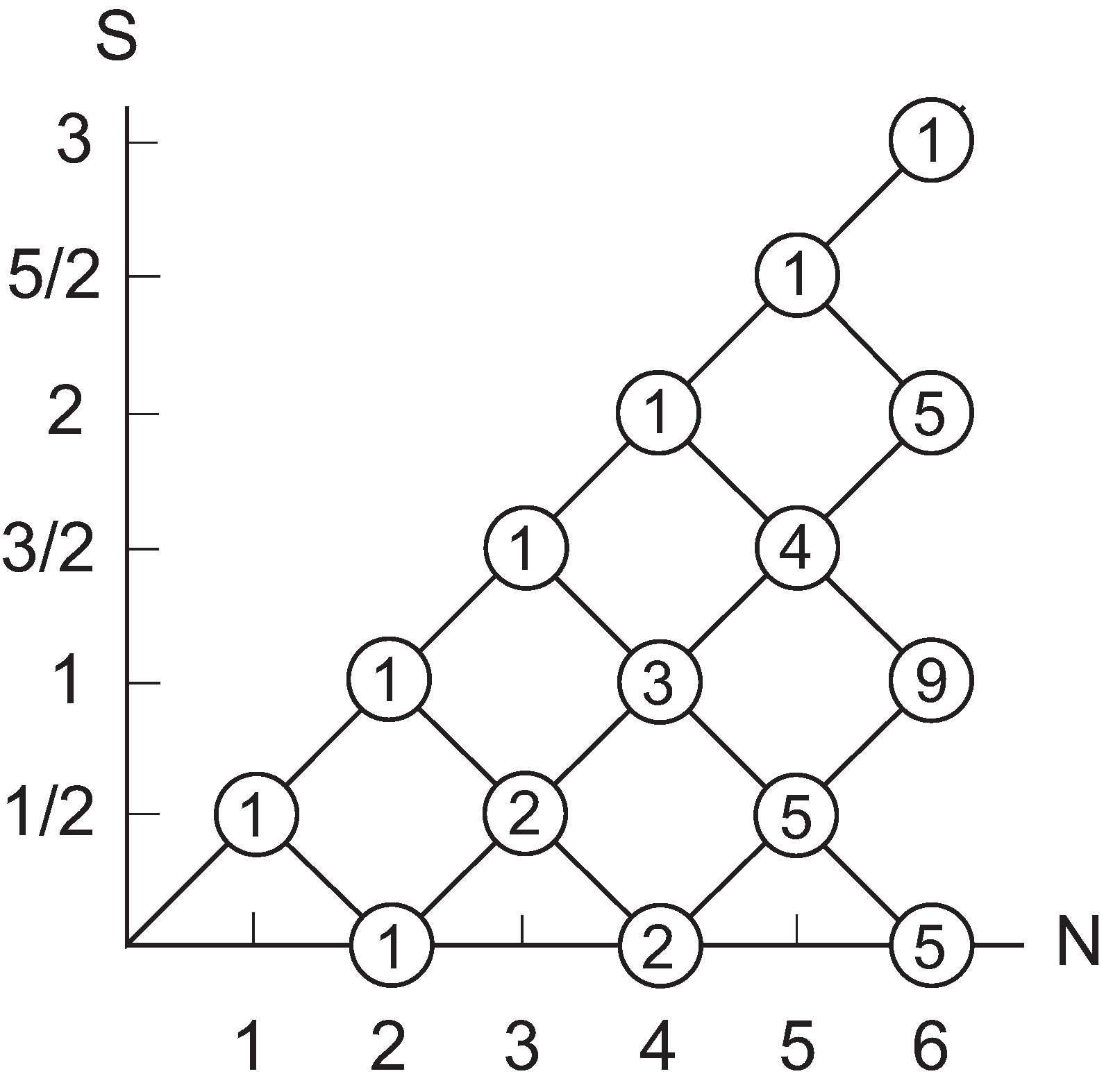}
\caption{The total-spin branching diagram illustrating the multiplicities 
${\cal G}(N,S)$ of the spin egenfunctions of $N$ spin-$1/2$ fermions. $S$ is the total spin.  
}
\label{brdi}
\end{figure}
We outline in this Appendix several properties of the many-body 
spin eigenfunctions which are useful for analyzing the trends and behavior
of the spin multiplicities exhibited by the CI wave functions for $N=4$ and $N=3$ 
ultracold fermions. The spin multiplicities of the CI wave functions lead naturally 
to analogies with finite Heisenberg clusters \cite{hend933,haas088} and to $t$-$J$-type
models.

A basic property of spin eigenfunctions is that they exhibit degeneracies 
for $N>2$, i.e., there may be more than one linearly independent (and 
orthogonal) spin functions that are simultaneous eigenstates of both 
$\hat{\bf S}^2$ and $S_z$. These degeneracies are usually visualized by 
means of the {\it branching diagram\/} \cite{pauncz} displayed in 
Fig.\ \ref{brdi}. The axes in this plot describe the number $N$ of fermions  
(horizontal axis) and the quantum number $S$ of the total spin (vertical axis).
At each point $(N,S)$, a circle is drawn containing the number ${\cal G}(N,S)$ which
gives the degeneracy of spin states. It is found \cite{pauncz} that
\begin{equation}
{\cal G}(N,S) = 
\left( \begin{array}{c} N \\ N/2-S \end{array} \right) -
\left( \begin{array}{c} N \\ N/2-S-1 \end{array} \right).
\label{degen}
\end{equation}

Specifically for $N=4$ particles, there is one spin eigenfunction with
$S=2$, three with $S=1$, and two with $S=0$. In general the spin part of the
CI wave functions involves a linear superposition over all the degenerate spin 
eigenfunctions for a given $S$. 

For a small number of particles, one can find compact expressions that encompass 
all possible superpositions. For example, for $N=4$ and $S=0$, $S_z=0$ one has: \cite{yann09,yingphd}

\begin{eqnarray}
 {\cal X}_{00}&=&\sqrt{\frac{1}{3}}\sin\theta \; 
   \alpha \alpha \beta \beta + \left( \frac{1}{2}\cos\theta- 
       \sqrt{\frac{1}{12}}\sin\theta \right)
   \alpha \beta \alpha \beta \nonumber \\
       & & - \left( \frac{1}{2}\cos\theta+\sqrt{\frac{1}{12}}\sin\theta \right)
   \alpha \beta \beta \alpha \nonumber \\
       & & -\left( \frac{1}{2}\cos\theta+\sqrt{\frac{1}{12}}\sin\theta \right)
   \beta \alpha \alpha \beta \nonumber \\
       & & + \left (\frac{1}{2}\cos\theta-\sqrt{\frac{1}{12}}\sin\theta \right)
   \beta \alpha \beta \alpha 
        +\sqrt{\frac{1}{3}}\sin\theta \;   
   \beta \beta \alpha \alpha, \nonumber \\
\label{x00}
\end{eqnarray}
where the parameter $\theta$ satisfies $-\pi/2\leq \theta \leq \pi/2$ 
and is chosen such that $\theta=0$ corresponds to the spin function with 
intermediate two-fermion spin $S_{12}=0$ and three-fermion spin
$S_{123}=1/2$; whereas $\theta=\pm \pi/2$ corresponds to the one with intermediate 
spins $S_{12}=1$ and $S_{123}=1/2$.

For $N=3$ and $S=1/2$, $S_z=1/2$ one has \cite{yingphd}:
\begin{eqnarray}
 {\cal X}_{1/2,1/2} &=& \sqrt{ \frac{2}{3} }\sin\theta \;\; \alpha \alpha \beta \nonumber \\
&+& \left( \sqrt{ \frac{1}{2} } \cos\theta -  \sqrt{ \frac{1}{6} } \sin\theta \right) \alpha \beta \alpha \nonumber \\
&-& \left( \sqrt{ \frac{1}{2} } \cos\theta +  \sqrt{ \frac{1}{6} } \sin\theta \right) \beta \alpha \alpha. 
\label{x1212}
\end{eqnarray}
For the general expressions for the remaining spin combinations, $S$ and $S_z$, for $N=4$ and $N=3$ fermions, 
see Refs.\ \cite{yann09,yingphd,vargabook}.

For each SPD corresponding to a given CI state of the system, one can plot four different spin-resolved CPDs, 
i.e., ${\cal P}_{\uparrow\uparrow}$, ${\cal P}_{\uparrow\downarrow}$,
${\cal P}_{\downarrow\uparrow}$, and ${\cal P}_{\downarrow\downarrow}$.
This can potentially lead to a very large number of time consuming computations and an excessive number of plots. 
For studying the spin structure of the $S=0, S_z=0$ states for $N=4$ fermions and the $S=1/2, S_z=1/2$ states for 
$N=3$, however, we found that knowledge of a single CPD, is sufficient in the regime of Wigner-molecule formation.
Indeed, the specific angle $\theta$ specifying the spin function ${\cal X}_{00}$ [Eq.\ (\ref{x00})] for $N=4$ or
the spin function ${\cal X}_{1/2,1/2}$ [Eq.\ (\ref{x1212})] for $N=3$ fermions can be determined through a
procedure exemplified in the following through two examples for $N=4$ fermions:

{\bf Example 1; case of the CPD in Fig.\ \ref{specpa}(g)}: 
The same labeling that numbers the sites determines also the left-to-right ordering of the localized electrons in 
each of the six primitive spin functions $\alpha \alpha \beta \beta$, $\alpha \beta \alpha \beta$, etc., that span
the eigenfunction ${\cal X}_{00}$ in Eq.\ (\ref{x00}). Namely, the fermion localized at the hump No. 1 corresponds
to the far left position in the primitive, the fermion localized at the hump No. 2 corresponds to the second from
the left position in the primitive, the fermion localized at the hump No. 3 corresponds to the third from the
left position in the primitive, and the fermion localized at the hump No. 4 corresponds to the far right position
in the primitive. The numbering of the humps does not necessarily follow the cardinal ordering 1,2,3,4, as will
become evident below from the second example concerning a linear Heisenberg chain.
An inspection of Eq.\ (\ref{x00}) shows that only the 
first three primitive spin functions in ${\cal X}_{00}$ can be associated with 
${\cal P}_{\downarrow\uparrow}({\bf r},{\bf r}_0 \equiv {\rm site~No.~1})$ [compare the CPD in Fig.\ 
\ref{specpa}(g)], namely $\alpha \alpha \beta \beta$, $\alpha \beta \alpha \beta$, and $\alpha \beta \beta \alpha$; 
these are the only primitives in Eq.\ (\ref{x00}) with a down spin in the site labeled as 1 [see diagram in Fig.\ 
\ref{specpa}(III)]. From these three primitives, only the first and the second contribute to the {\it partial\/} 
conditional probability $\Pi_{\downarrow\uparrow}(4,1)$ of finding another fermion with spin-down in site No. 4,
while the first fermion is fixed at site No. 1. 
Taking the squares of the coefficients of $ \alpha \alpha \beta \beta$ and 
$\alpha \beta \alpha \beta$ in Eq.\ (\ref{x00}), one gets
\begin{equation}
\Pi_{\downarrow\uparrow}(4,1) 
\propto 
\frac{\sin^2 \theta}{3} +  
\left( \frac{1}{2} \cos \theta - \sqrt{ \frac{1}{12} } \sin \theta \right)^2.
\label{pi_downup_41}
\end{equation} 
Similarly, one finds
\begin{eqnarray}
\Pi_{\downarrow\uparrow}(2,1) &\propto&
\left( \frac{1}{2}\cos\theta-\sqrt{\frac{1}{12}}\sin\theta \right)^2 \nonumber \\
&+& \left( \frac{1}{2}\cos\theta+\sqrt{\frac{1}{12}}\sin\theta \right)^2
\label{pi_downup_21}
\end{eqnarray}
and
\begin{equation}
\Pi_{\downarrow\uparrow}(3,1) \propto
\frac{\sin^2 \theta}{3} +  
\left( \frac{1}{2}\cos\theta+\sqrt{\frac{1}{12}}\sin\theta \right)^2
\label{pi_downup_31}
\end{equation}

The quantities $\Pi_{\downarrow\uparrow}(i,1)$, $i=2,3,4$, as defined above correspond to the volumes 
Vol$(i)$, $i=2,3,4$ under the humps labeled No. 2, No. 3, and No. 4 of the CI CPD in Fig.\ \ref{specpa}(g).  
Integrating numerically under the humps of the CI CPD in Fig.\ \ref{specpa}(g), we specify the ratio
$x=$Vol(4)/[Vol(2)+Vol(3)], which yields the condition
\begin{equation}
\frac{\Pi_{\downarrow\uparrow}(4,1)}{\Pi_{\downarrow\uparrow}(2,1)+\Pi_{\downarrow\uparrow}(3,1)}=x,
\label{piratio}
\end{equation}
For the case of Fig.\ \ref{specpa}(g), we find $x=1$. For $x=1$, condition (\ref{piratio}) can be satisfied for 
an angle $\theta=-\pi/3$ [compare with the spin eigenfunction in Eq.\ (\ref{x00_cpd1})].

{\bf Example 2; case of the CPD in Fig.\ \ref{specpa}(c)}:
As a second example, we choose the case of a single well. Illustrative calculations for the spectrum,
densities, and CPDs for this case are displayed in Fig.\ \ref{specpa}(a,b,c). Note the labeling of the four sites in 
space, which is ``4123'' and not ``1234''. This results from our taking $J_{34}=0$, when opening the four-site ring,
and it is consistent with our treatment of the four-site linear Heisenberg chain in \ref{ac} below.   

Noting that hump No. 4 in Fig.\ \ref{specpa}(c) is again well isolated from the rest, and 
focussing on the numbering of the remaining humps of this SR-CPD, it is apparent that we
need to use the same set of the quantities $\Pi_{\downarrow\uparrow}(i,1)$, $i=2,3,4$ as was the case with
the previous example. Integrating under the humps of the CI CPD in Fig.\ \ref{specpa}(c), we find the numerical
values for the volumes Vol$(i)$, $i=2,3,4$. In particular, we determine that $x=0.789$. With this value of the 
ratio $x$, condition (\ref{piratio}) yields an angle of $\theta=-\pi/5.12$.

{\bf Example 3; case of the CPD in Fig.\ \ref{specpa}(i)}:
As a third example, we choose an excited state (the one with $S=0$ and $S_z=0$) in the double well. Illustrative 
calculations for the spectrum, densities, and CPDs for this case are displayed in Fig.\ \ref{specpa}(e,h,i). Note 
again the labeling of the four sites in space. Noting that hump No. 4 in Fig.\ \ref{specpa}(i) is again well isolated 
from the rest, and focussing on the numbering of the remaining humps of this SR-CPD, it is apparent that we
need to use the same set of the quantities $\Pi_{\downarrow\uparrow}(i,1)$, $i=2,3,4$ as was the case with
the previous examples. Integrating under the humps of the CI CPD in Fig.\ \ref{specpa}(i), we find the numerical
values for the volumes Vol$(i)$, $i=2,3,4$. In particular, we determine that $x=0.20$ in this case. With this value of
the ratio $x$, condition (\ref{piratio}) yields an angle of $\theta=\pi/6$.

For further detailed applications of this procedure, see Refs.\ \cite{yann09,yingphd}.

\section{Heisenberg model for 4 localized fermions in a DWPA configuration}
\label{ab}

\begin{figure}[t]
\centering\includegraphics[width=5cm]{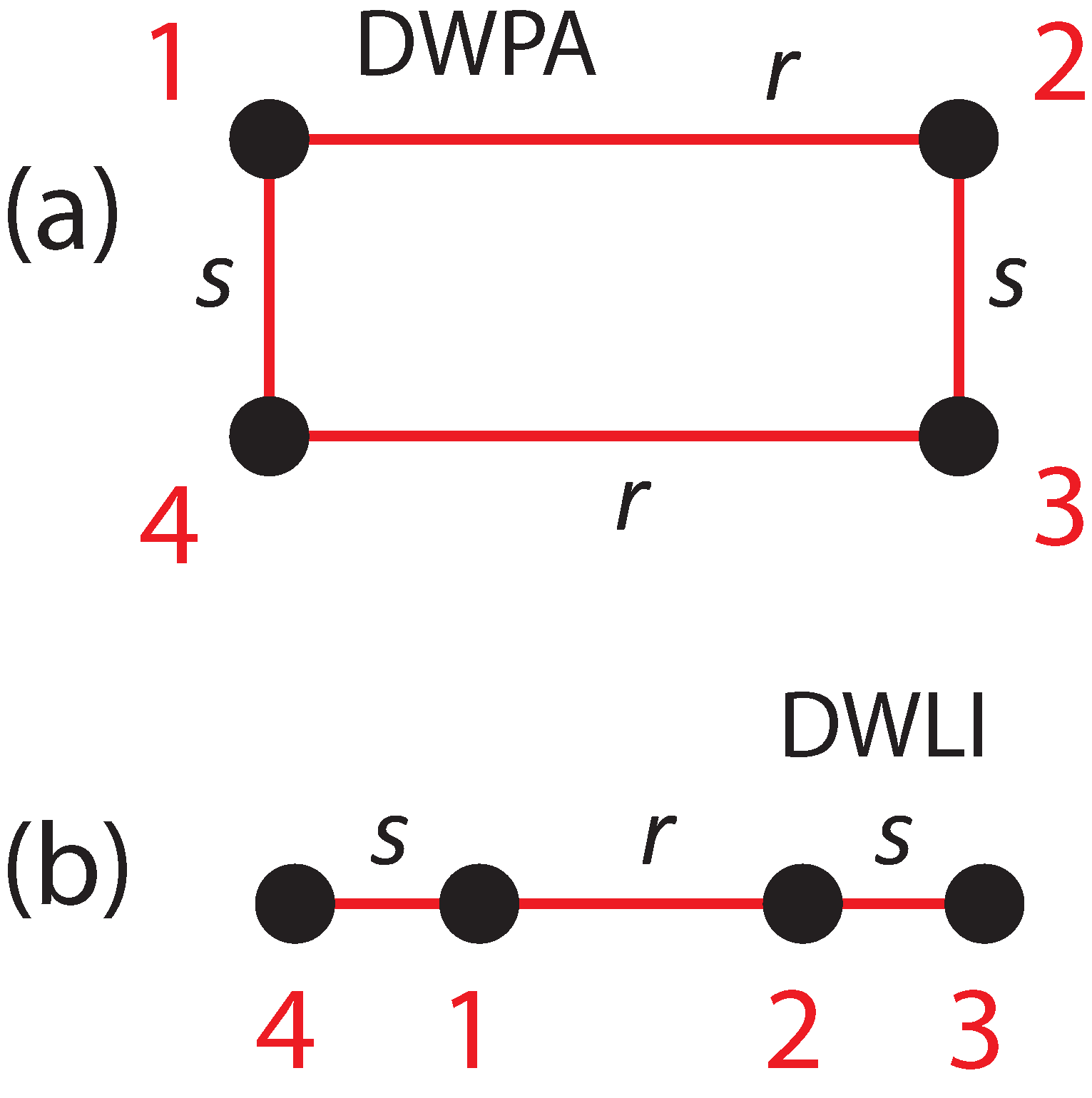}
\caption{Schematics indicating the four-site numbering convention in the Heisenberg Hamiltonian.
(a) The case of formation of a rectangular parallelogram (ring topology). The Heisenberg exchange parameters 
$J_{14}=J_{23}=s$ and $J_{12}=J_{34}=r$. (b) The linear arrangement of the four sites which results from (a) by 
opening the ring through setting $J_{34}=0$, $J_{12}=r$, $J_{14}=J_{23}=s$.
}
\label{rect}
\end{figure}

The single particle densities and CPDs in Figs. \ref{specpa} and \ref{specli} show that the associated 
Wigner-molecule CI wave functions can be mapped onto the spin functions for four fermions. These spin functions are 
solutions of a 4-site Heisenberg Hamiltonian ${\cal H}_H^{\rm RP,gen}$ with the four fermions being located at the 
vertices of a rectangular parallelogram (RP) in the case of the double-well parallel arrangement. Assuming for the 
sake of generality that all nearest-neighbor exchange couplings $J_{ij}$ are different, one has 
\begin{equation}
\fl
{\cal H}_H^{\rm RP,gen} = J_{12} {\bf S}_1 {\bf \cdot S}_2 + J_{23} {\bf S}_2 {\bf \cdot S}_3 + 
J_{34} {\bf S}_3 {\bf \cdot S}_4 + J_{14} {\bf S}_1 {\bf \cdot S}_4 -(J_{12}+J_{23}+J_{34}+J_{14})/4,
\label{hh1}
\end{equation}
where the indices $k$ in ${\bf S}_k$ denote the locations of the four sites, which are associated 
with the four humps in the s.p. density of Fig.\ \ref{specpa} (in a clockwise direction); see also 
schematic in Fig.\ \ref{rect}(a). 

For the case of all four fermions being trapped in a single well, one has an open linear 4-site Heisenberg chain, 
which is obtained from Eq.\ (\ref{hh1}) by setting $J_{34}=0$, i.e.,
\begin{equation}
\fl
{\cal H}_H^{\rm 4LI,gen} = J_{12} {\bf S}_1 {\bf \cdot S}_2 + J_{23} {\bf S}_2 {\bf \cdot S}_3 +
J_{14} {\bf S}_1 {\bf \cdot S}_4 -(J_{12}+J_{23}+J_{14})/4,
\label{hh2}
\end{equation}

To proceed, it is sufficient to use the six-dimensional Ising 
subspace for zero total-spin projection ($S_z=0$), which is spanned by the 
following set of basis states: 
$|1\rangle \rightarrow  \alpha\alpha\beta\beta$,
$|2\rangle \rightarrow  \alpha\beta\alpha\beta$, 
$|3\rangle \rightarrow  \alpha\beta\beta\alpha$,
$|4\rangle \rightarrow  \beta\alpha\alpha\beta$,
$|5\rangle \rightarrow  \beta\alpha\beta\alpha$, and
$|6\rangle \rightarrow  \beta\beta\alpha\alpha$;
the ordering from left to right coincides with the cardinal ordering 
$1,\ldots,4$ of the sites in Figs.\ \ref{rect}(a) and \ref{rect}(b).

Using the raising and lowering operators $S_i^+=S_i^x + i S_i^y$, $S_i^-=S_i^x - i S_i^y$,
and the identity ${\bf S}_i {\bf \cdot S}_j=S_i^x S_j^x + S_i^y S_j^y + S_i^z S_j^z$, the
Heisenberg Hamiltonian given by Eq. (\ref{hh1}) can be written as
\begin{eqnarray}
{\cal H}_H^{\rm RP,gen} 
&=& J_{12} S_1^z S_2^z + J_{14} S_1^z S_4^z + J_{23} S_2^z S_3^z + J_{34} S_3^z S_4^z \nonumber \\
&+& \frac{J_{12}}{2} S_1^+ S_2^- + \frac{J_{14}}{2} S_1^+ S_4^- 
+ \frac{J_{23}}{2} S_2^+ S_3^- + \frac{J_{34}}{2} S_3^+ S_4^-) \nonumber \\
&+& \frac{J_{12}}{2} S_1^- S_2^+ + \frac{J_{14}}{2} S_1^- S_4^+
+ \frac{J_{23}}{2} S_2^- S_3^+ + \frac{J_{34}}{2} S_3^- S_4^+ \nonumber \\
&-& (J_{12}+J_{23}+J_{34}+J_{14})/4.
\label{hh1int}
\end{eqnarray}

With the relations $S_i^z \alpha =\alpha/2$, $S_i^z \beta = -\beta/2$, $S_i^+ \alpha=0$, $S_i^+ \beta=\alpha$,
$S_i^- \alpha=\beta$, $S_i^- \beta=0$, one can write ${\cal H}_H^{\rm RP,gen}$ in matrix form, as follows 
{\scriptsize
\begin{eqnarray} 
&& \fl {\cal H}_H^{\rm RP,gen} = \nonumber \\
&& \fl \frac{1}{2} \left(
\begin{array}{cccccc} 
-J_{14}-J_{23}  &  J_{23}  &  0  & 
0  &  J_{14}  &  0  \\
J_{23} & -J_{12}-J_{14}-J_{23}-J_{34}  &  J_{34} & 
J_{12} & 0 & J_{14} \\
0 & J_{34} & -J_{12}-J_{34} & 
0 & J_{12} & 0  \\
0 & J_{12} & 0 & 
-J_{12}-J_{34} & J_{34} & 0 \\
J_{14} & 0 & J_{12} & 
J_{34} & -J_{12}-J_{14}-J_{23}-J_{34} & J_{23} \\
0 & J_{14} & 0 & 
0 & J_{23} & -J_{14}-J_{23} 
\end{array} 
\right) 
. \nonumber \\
\label{hh1matgen} 
\end{eqnarray} 
}

Due to the reflection symmetry in $x$ and $y$, ${\cal H}_H^{\rm RP,gen}$ has only two different exchange constants 
$J_{14}=J_{23}=s$ and $J_{12}=J_{34}=r$. ($r$ here decreases rapidly with the distance, or the interwell 
barrier height.) As a result, the matrix form of ${\cal H}_H^{\rm RP,gen}$ simplifies to the following
\begin{equation} 
\fl {\cal H}_H^{\rm RP} = \left(
\begin{array}{cccccc} 
-s & s/2 & 0 & 0 & s/2 & 0  \\
s/2 & - r - s & r/2 & r/2 & 0 & s/2 \\
0 & r/2 & -r & 0 & r/2 & 0  \\
0 & r/2 & 0 & -r & r/2 & 0 \\
s/2 & 0 & r/2 & r/2 & - r - s & s/2 \\
0 & s/2 & 0 & 0 & s/2 & -s 
\end{array} 
\right) \;\;
\begin{array}{c}
\alpha \alpha \beta \beta \\
\alpha \beta \alpha \beta \\
\alpha \beta \beta \alpha \\
\beta \alpha \alpha \beta \\
\beta \alpha \beta \alpha \\
\beta \beta \alpha \alpha 
\end{array}
.
\label{hh1mat} 
\end{equation} 

The general eigenvalues ${\cal E}_i$ and corresponding eigenvectors ${\cal V}_i$ of the matrix (\ref{hh1mat})
are calculated easily using MATHEMATICA \cite{mathsm}. The eigenvalues are:
\begin{equation}
{\cal E}_1=- s - r,\;\;S=1,
\label{e1}
\end{equation}
\begin{equation}
{\cal E}_2=-r,\;\;S=1,
\label{e2}
\end{equation}
\begin{equation}
{\cal E}_3=-s,\;\;S=1,
\label{e3}
\end{equation}
\begin{equation}
{\cal E}_4=0,\;\;S=2,
\label{e4}
\end{equation}
\begin{equation}
{\cal E}_5= - s - r - {\cal Q}(s,r),\;\;S=0,
\label{e5}
\end{equation}
\begin{equation}
{\cal E}_6= - s - r + {\cal Q}(s,r),\;\;S=0,
\label{e6}
\end{equation}
where 
\begin{equation}
{\cal Q}(a,b)=\sqrt{a^2-a b+b^2}.  
\label{qq}
\end{equation}

The corresponding (unnormalized) eigenvectors and their total spins are given by:
\begin{equation}
{\cal V}_1=
\{0, -1, 0, 0, 1, 0 \}^T,\;\;S=1,
\label{v1}
\end{equation}
\begin{equation}
{\cal V}_2=
\{0, 0, -1, 1, 0, 0 \}^T,\;\;S=1,
\label{v2}
\end{equation}
\begin{equation}
{\cal V}_3=
\{ -1, 0, 0, 0, 0, 1 \}^T,\;\;S=1,
\label{v3}
\end{equation}
\begin{equation}
{\cal V}_4=
\{1, 1, 1, 1, 1, 1 \}^T,\;\;S=2,
\label{v4}
\end{equation}
\begin{equation}
{\cal V}_5=
\{1, -{\cal X}, -1+{\cal X}, -1+{\cal X}, -{\cal X}, 1 \}^T,\;\; S=0,
\label{v5}
\end{equation}
\begin{equation}
{\cal V}_6=
\{1, -{\cal Y}, -1+{\cal Y}, -1+{\cal Y}, -{\cal Y}, 1 \}^T,\;\; S=0,
\label{v6}
\end{equation}
where 
\begin{equation}
{\cal X}=f+{\cal Q}(1,f),
\label{xq}
\end{equation}
\begin{equation}
{\cal Y}=f-{\cal Q}(1,f),
\label{yq}
\end{equation}
and $f=r/s$.

To understand how the Heisenberg Hamiltonian in Eq.\ (\ref{hh1mat}) captures the behavior seen in the 
CI spectra of Fig.\ \ref{specpa} (DWPA case), we start with the limiting case $r \rightarrow 0$, which is 
applicable (see below) to the larger interwell barrier $V_b= 11.14$ kHz. In this limit,
one can neglect $r$ compared with $s$, which results in a characteristic 1-2-3 degeneracy pattern 
within the band; namely one has ${\cal E}_2={\cal E}_4={\cal E}_6=0$, ${\cal E}_1={\cal E}_3=-s$, and ${\cal E}_5=-2s$.

Furthermore, the fact that all six curves in the CI lowest-energy 
band cross at the same point $1/g=0$  suggests that $s \sim F(-1/g)$ and $r \sim F(-1/g)$ with ($x=-1/g$)
\begin{equation} 
F(x)=\tanh(x). 
\label{f14}
\end{equation}

Of interest is the fact that the ability of the Heisenberg Hamiltonian
in Eq.\ (\ref{hh1mat}) to reproduce the CI trends is not restricted solely to
energy spectra, but extends to the CI wave functions as well. Indeed when
$r \rightarrow 0$, the last two eigenvectors of the Heisenberg 
matrix (having $S=0$) become
\begin{equation}
{\cal V}_5 \rightarrow \{1,-1,0,0,-1,1\}^T,
\label{v52}
\end{equation}
and
\begin{equation}
{\cal V}_6 \rightarrow \{1,1,-2,-2,1,1\}^T.
\label{v62}
\end{equation}
When multiplied by the normalization factor, the wave functions represented by 
the eigenvectors in Eq.\ (\ref{v52}) coincides (within an overall $\mp 1$ sign) with 
the ground-state CI spin function ${\cal X}_{00}^{(1)}$ in Eq.\ (\ref{x00_cpd1}). 

The CI spectra and spin functions for the smaller barrier  $V_b = 6.08$ kHz can
be analyzed within the framework of the 4-site Heisenberg Hamiltonian
(\ref{hh1mat}) when small (compared with $J_{14}=s$), but nonnegligible, 
values of the second exchange integral $J_{12}=r$ are considered. In this 
case, the partial three-fold and two-fold degeneracies are lifted. Indeed in 
Figs.\ \ref{specpa}(d)  ($V_b=6.08$ kHz), the CI lowest-energy band consists of six distinct levels. 

\section{Heisenberg model for 4 localized fermions in a DWLI configuration}
\label{ac}

In the case of a single well and of a double-well in a linear arrangement, the Heisenberg Hamiltonian 
${\cal H}_H^{\rm 4LI,gen}$ in Eq.\ (\ref{hh2}) is of relevance. In the Ising basis, and using $J_{14}=J_{23}=s$, 
$J_{12}=r$, and $J_{34}=0$ in Eq.\ (\ref{hh1matgen}), this Hamiltonian reduces to ${\cal H}_H^{\rm 4LI}$, i.e.,
\begin{equation}
\fl {\cal H}_H^{\rm 4LI} = \left(
\begin{array}{cccccc}
- s & s/2 & 0 & 0 & s/2 & 0  \\
s/2 & - (r/2 + s) & 0 & r/2 & 0 & s/2 \\
0 & 0 & - r/2 & 0 & r/2 & 0  \\
0 & r/2 & 0 & - r/2 & 0 & 0 \\
s/2 & 0 & r/2 & 0 & - (r/2 + s) & s/2 \\
0 & s/2 & 0 & 0 & s/2 & - s
\end{array} 
\right) \;\;
\begin{array}{c}
\alpha \alpha \beta \beta \\
\alpha \beta \alpha \beta \\
\alpha \beta \beta \alpha \\
\beta \alpha \alpha \beta \\
\beta \alpha \beta \alpha \\
\beta \beta \alpha \alpha 
\end{array}
.
\label{hh2mat}
\end{equation}

The general eigenvalues of the matrix (\ref{hh2mat}) are:
\begin{equation}
{\cal E}_1 = -s,\;\; S=1,
\label{e2_1}
\end{equation}

\begin{equation}
{\cal E}_2 = (-r-s+\sqrt{r^2+s^2})/2,\;\; S=1, 
\label{e2_2}
\end{equation}

\begin{equation}
{\cal E}_3 = (-r-s-\sqrt{r^2+s^2})/2,\;\; S=1, 
\label{e2_3}
\end{equation}

\begin{equation}
{\cal E}_4 = 0,\;\; S=2,
\label{e2_4}
\end{equation}

\begin{equation}
{\cal E}_5 = -r/2-s-{\cal Q}(2s,r)/2,\;\; S=0,
\label{e2_5}
\end{equation}

\begin{equation}
{\cal E}_6 = -r/2-s+{\cal Q}(2s,r)/2,\;\; S=0.
\label{e2_6}
\end{equation}

The corresponding (unnormalized) eigenvectors and their total spins are given by:
\begin{equation}
\fl {\cal V}_1=
\{-1, 0, 0, 0, 0, 1 \}^T,\;\;S=1,
\label{v2_1}
\end{equation}
\begin{equation}
\fl {\cal V}_2=
\{0, -1, -r/(s-\sqrt{s^2+r^2}), r/(s-\sqrt{s^2+r^2}), 1, 0 \}^T,\;\;S=1,
\label{v2_2}
\end{equation}
\begin{equation}
\fl {\cal V}_3=
\{0, -1, -r/(s+\sqrt{s^2+r^2}), r/(s+\sqrt{s^2+r^2}), 1, 0 \}^T,\;\;S=1,
\label{v2_3}
\end{equation}
\begin{equation}
\fl {\cal V}_4=
\{1, 1, 1, 1, 1, 1 \}^T,\;\;S=2,
\label{v2_4}
\end{equation}
\begin{equation}
\fl {\cal V}_5=
\{ 1, -{\cal W}, -1+{\cal W}, -1+{\cal W}, -{\cal W}, 1 \}^T,\;\; S=0,
\label{v2_5}
\end{equation}
\begin{equation}
\fl {\cal V}_6=
\{ 1, -{\cal Z}, -1+{\cal Z}, -1+{\cal Z}, -{\cal Z}, 1 \}^T,\;\; S=0,
\label{v2_6}
\end{equation}
where 
\begin{equation}
{\cal W}=f/2+{\cal Q}(1,f/2),
\end{equation}

\begin{equation}
{\cal Z}=f/2-{\cal Q}(1,f/2),
\end{equation}
and $f=r/s$ as previously defined.

In the limit of $r \rightarrow 0$ (high interwell barrier $V_b$),
the energies in Eqs.\ (\ref{e2_1})-(\ref{e2_6}) reproduce the characteristic 1-2-3 degeneracy pattern, which
appears also in the case of the rectangular arrangement of the four sites; namely one 
has ${\cal E}_2={\cal E}_4={\cal E}_6=0$, ${\cal E}_1={\cal E}_3=-s$, and ${\cal E}_5=-2s$. Furthermore,
for $ r \rightarrow 0$, the corresponding (unnormalized) eigenvectors and their total spins are given by:
\begin{equation}
{\cal V}_1=
\{-1, 0, 0, 0, 0, 1 \}^T,\;\;S=1,
\label{vv2_1}
\end{equation}
\begin{equation}
{\cal V}_2=
\{0, 0, 1, -1, 0, 0 \}^T,\;\;S=1,
\label{vv2_2}
\end{equation}
\begin{equation}
{\cal V}_3=
\{ 0, -1, 0, 0, 1, 0 \}^T,\;\;S=1,
\label{vv2_3}
\end{equation}
\begin{equation}
{\cal V}_4=
\{1, 1, 1, 1, 1, 1 \}^T,\;\;S=2,
\label{vv2_4}
\end{equation}
\begin{equation}
{\cal V}_5=
\{ 1, -1, 0, 0, -1, 1 \}^T,\;\; S=0,
\label{vv2_5}
\end{equation}
\begin{equation}
{\cal V}_6=
\{1, 1, -2, -2, 1, 1 \}^T,\;\; S=0.
\label{vv2_6}
\end{equation}

Note that the open-chain eigenvectors (\ref{v2_1})-(\ref{v2_6}) coincide with those 
[see Eqs.\ (\ref{v1})-(\ref{v6})] of the closed-chain rectangular configuration when $r \rightarrow 0$.

From a fitting of the open-chain Heisenberg eigenvalues to the CI spectrum (see Sec.\ \ref{4fhh} above),
we found that the case of 4 fermions in a single quasi-1D harmonic trap is described well when $r=1.35s$.   
Indeed, in this case, all the eigenvalues are different. Specifically, with the value of $r/s=1.35$, the open linear 
Heisenberg chain yields 
${\cal E}_1=-s$ ($S=1$), ${\cal E}_2=-0.334985 s$ ($S=1$), ${\cal E}_3=-2.01501 s$ ($S=1$), 
${\cal E}_4=0$ ($S=2$),  ${\cal E}_5=-2.55853 s$ ($S=0$), and ${\cal E}_6=-0.79147 s$ ($S=0$),

The corresponding (normalized) Heisenberg eigenvectors are given by:
\begin{equation}
\fl {\cal V}_1=
\{1/\sqrt{2}, 0, 0, 0, 0, -1/\sqrt{2} \}^T,\;\;S=1,
\label{v3_1}
\end{equation}
\begin{equation}
\fl {\cal V}_2=
\{0, -0.318109, 0.631512, -0.631512, 0.318109, 0\}^T,\;\;S=1,
\label{v3_2}
\end{equation}
\begin{equation}
\fl {\cal V}_3=
\{0, -0.631512, -0.318109, 0.318109, 0.631512, 0\}^T,\;\; S=1,
\label{v3_3}
\end{equation}
\begin{equation}
\fl {\cal V}_4=
\{1/\sqrt{6}, 1/\sqrt{6}, 1/\sqrt{6}, 1/\sqrt{6}, 1/\sqrt{6}, 1/\sqrt{6} \}^T,\;\; S=2,
\label{v3_4}
\end{equation}
\begin{equation}
\fl {\cal V}_5=
\{0.365589, -0.569781, 0.204192, 0.204192, -0.569781, 0.365589 \}^T,\;\;S=0,
\label{v3_5}
\end{equation}
\begin{equation}
\fl {\cal V}_6=
\{0.446854, 0.0931823, -0.540036, -0.540036, 0.0931823, 0.446854\}^T,\;\;S=0.
\label{v3_6}
\end{equation}

\begin{figure}[t]
\centering\includegraphics[width=5cm]{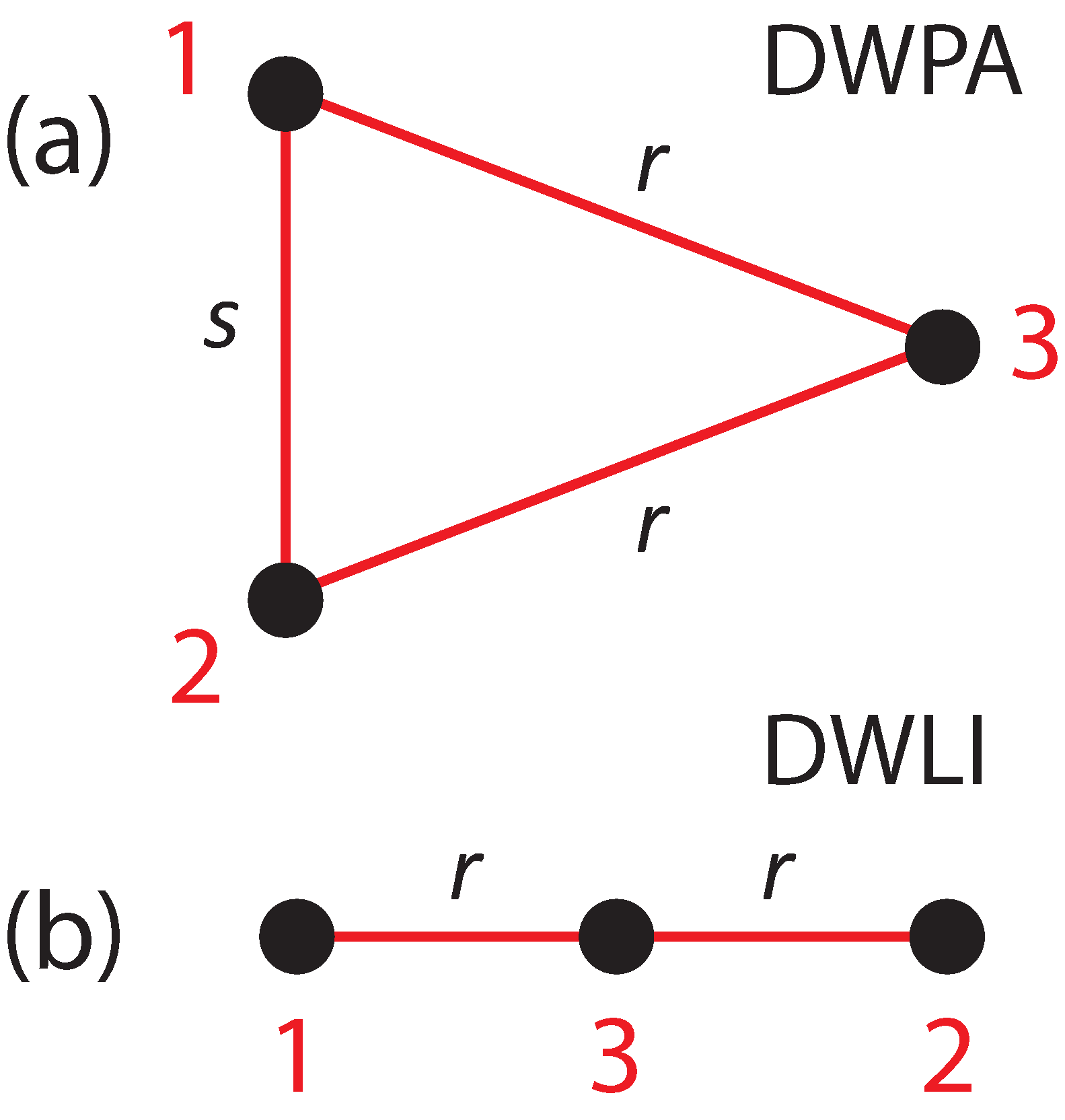}
\caption{Schematics of the three-site numbering conventions in the Heisenberg Hamiltonian.
(a) The case of formation of an isosceles triangle and (b) a linear arrangement of the sites. 
$s$ and $r$ denote Heisenberg exchange parameters.
}
\label{trg}
\end{figure}
~~~~~\\
~~~~~\\

\section{Heisenberg model for 3 localized fermions in tilted wells}
\label{heis3}

In the case of $N=3$ strongly-interacting fermions in a single 1D well or a tilted double-well with a parallel 
arrangement of the two 1D wells, the simple Heisenberg model is applicable. For a $(2,1)$ fermion configuration, the
three sites form an isosceles triangle (see Figs.\ \ref{specn3} and \ref{trg}), and the associated 
Heisenberg-ring Hamiltonian ${\cal H}_H^{\rm trg}$ is given by Eq.\ (\ref{heihtrg}).
To proceed, we use the three-dimensional Ising Hilbert subspace for total-spin projection 
$S_z=1/2$, which is spanned by the following set of basis states:
$|1> \rightarrow \alpha \alpha \beta$, 
$|2> \rightarrow \alpha \beta \alpha$, and
$|3> \rightarrow \beta \alpha \alpha$.
In this subspace, the complete Heisenberg Hamiltonian in Eq. (\ref{heihtrg}) can be 
written in matrix form as ($J_{12}=s$, $J_{13}=J_{23}=r$)
\begin{equation} 
{\cal H}_H^{{\rm trg}} = \left(
\begin{array}{ccc} 
-r & r/2 & r/2 \\
r/2 & -s/2-r/2 & s/2 \\ 
r/2 & s/2 & -s/2-r/2 
\end{array} 
\right)\;\;
\begin{array}{l}
\alpha \alpha \beta \\
\alpha \beta \alpha \\
\beta \alpha \alpha \\
\end{array}
.
\label{hh3mat} 
\end{equation} 

The general eigenvalues of the matrix (\ref{hh3mat}) are:
\begin{equation}
{\cal E}_1 = 0,\;\; S=3/2,
\label{e3_1}
\end{equation}

\begin{equation}
{\cal E}_2 = -3r/2,\;\; S=1/2, 
\label{e3_2}
\end{equation}

\begin{equation}
{\cal E}_3 = -s-r/2,\;\; S=1/2. 
\label{e3_3}
\end{equation}

The corresponding (unnormalized) eigenvectors and their total spins are given by:
\begin{equation}
{\cal V}_1=
\{1, 1, 1 \}^T,\;\;S=3/2,
\label{v33_1}
\end{equation}
\begin{equation}
{\cal V}_2=
\{-2, 1, 1 \}^T,\;\;S=1/2,
\label{v33_2}
\end{equation}
\begin{equation}
{\cal V}_3=
\{ 0, -1, 1 \}^T,\;\;S=1/2,
\label{v33_3}
\end{equation}

Note that the eigenvectors are independent of $s$ and $r$, however which one is the ground state depends 
on these exchange constants through the expressions for the eigenvalues ${\cal E}_i$ given in 
Eqs.\ (\ref{e3_1})-(\ref{e3_3}). In particular, when the interwell barrier is high ($r \rightarrow 0$) 
[see Fig.\ \ref{specn3}(j)] a characteristic 1-2 degeneracy develops with ${\cal E}_1={\cal E}_2=0$ and
${\cal E}_3=-s$. When $s>0$, the ground-state vector is given by ${\cal V}_3$ in Eq.\ (\ref{v33_3}).

The case of 3 fermions in a single well [forming a linear Wigner molecule, see Fig.\ \ref{specn3}(VI)] is
described by the matrix Hamiltonian (\ref{hh3mat}) when $s=0$ (open Heisenberg chain). Then all three eigenvalues 
are different with ${\cal E}_1=0$, ${\cal E}_2=-3r/2$, and ${\cal E}_3=-r/2$ [see Fig.\ \ref{specn3}(n)]. 
Thus, with $r>0$, the ground-state vector for the (3,0) fermion arrangement is given by ${\cal V}_2$ in Eq.\ 
(\ref{v33_2}) and is different from that of the (2,1) fermion arrangement, although the total spin remains the same,
i.e., $S=1/2$ [compare SR-CPDs in Figs.\ \ref{specn3}(p,l)].

\section{The $t$-$J$ model for 3 localized fermions in a symmetric double well}
\label{tj3}

The (2,1) case of three strongly-interacting fermions in a tilted double well [with $\Delta=0.5$ kHz, see
Fig.\ \ref{specn3}(k)] is associated with a single triangular Wigner molecule. However, a more
complex WM configuration emerges when $\Delta =0$, i.e., for a symmetric double well. A remarkable
manifestation of this complexity is the doubling (from three to six) of the curves comprising the lowest energy 
band [contrast Fig.\ \ref{specn3}(j) and Fig.\ \ref{specn3}(a,f)]. This doubling of the energy curves 
indicates the presence of two resonating underlying configurations. Indeed, in order to satisfy parity conservation,
the single (135) triangle (see diagram in Fig.\ \ref{2trg}) needs to be supplemented with its mirror
configuration (246). This points to a model with 6 crystal sites, where 3 of them are occupied while
the remaining 3 are empty. This results in two Heisenberg clusters that are coupled via the tunneling (coherent 
hopping, with matrix elements denoted as $t_j$) of the fermions between the two triangular configurations (135) and 
(246). Since each of the six sites can assume three values, spin-up ($\alpha$), spin-down ($\beta$), and 
empty (0), one needs to use a generalization of the Ising Hilbert space spanned by the basis:
$|1> \rightarrow \alpha 0 \alpha 0 \beta  0$,   
$|2> \rightarrow \alpha 0 \beta  0 \alpha 0$,   
$|3> \rightarrow \beta  0 \alpha 0 \alpha 0$,   
$|4> \rightarrow 0 \beta 0 \alpha 0 \alpha $,   
$|5> \rightarrow 0 \alpha 0 \alpha 0 \beta $, and   
$|6> \rightarrow 0 \alpha 0 \beta 0 \alpha $.   

We have found that the CI results in Figs.\ \ref{specn3}(a) and \ref{specn3}(f) can be reproduced by using
two hopping parameters only, i.e., by setting $t_1=t_3=t$; for the definition of $t_j$'s, see the main text.
In this case, the relevant generalization of the Heisenberg Hamiltonian matrix in Eq.\ (\ref{hh3mat}) is given by\\
~~~~~~~~\\
{\small
\begin{equation}
\fl {\cal H}_{tJ}^{\Delta=0} = \left(
\begin{array}{ccc|ccc}
-r      & r/2        & r/2       & t_2      & t         & t         \\
r/2     & -s/2-r/2   & s/2       & t        & t         & t_2       \\
r/2     & s/2        & -s/2-r/2  & t        & t_2       & t         \\ \hline
t_2     & t          & t         & -r       & r/2       & r/2       \\
t       & t          & t_2       & r/2      & -s/2-r/2  & s/2       \\
t       & t_2        & t         & r/2      & s/2       & -s/2-r/2  \\
\end{array}
\right)\;\;\;
\begin{array}{l}
\alpha 0 \alpha 0 \beta  0  \\
\alpha 0 \beta  0 \alpha 0  \\
\beta  0 \alpha 0 \alpha 0  \\
0 \beta 0 \alpha 0 \alpha   \\
0 \alpha 0 \alpha 0 \beta   \\
0 \alpha 0 \beta 0 \alpha   \\
\end{array}
.
\label{tjmat}
\end{equation}
}
~~~~~~~\\

The $t$-$J$ Hamiltonian matrix in Eq.\ (\ref{tjmat}) exhibits a very rich behavior. In the following, we will limit 
our analysis to the case with $r=0$, i.e., for large interwell barrier $V_b$ which is also the case of both spectra
in Fig.\ \ref{specn3}(a) and \ref{specn3}(f). In this limit, the eigenvalues and eigenvectors (unnormalized) of the 
Hamiltonian matrix (\ref{tjmat}) are:
\begin{equation}
{\cal E}_1=-s + t - t_2,\;\; S=1/2,
\label{e4_1}
\end{equation}
\begin{equation}
{\cal E}_2=-s - t + t_2,\;\; S=1/2,
\label{4_2}
\end{equation}
\begin{equation}
{\cal E}_3=t - t_2,\;\; S=1/2,
\label{e4_3}
\end{equation}
\begin{equation}
{\cal E}_4=-t + t_2,\;\; S=1/2,
\label{e4_4}
\end{equation}
\begin{equation}
{\cal E}_5=2 t + t_2,\;\; S=3/2,
\label{e4_5}
\end{equation}
\begin{equation}
{\cal E}_6=-2 t - t_2,\;\; S=3/2,
\label{e4_6}
\end{equation}

and
\begin{equation}
{\cal V}_1=
\{0, -1, 1, 0, -1, 1 \}^T,\;\;S=1/2,
\label{v4_1}
\end{equation}
\begin{equation}
{\cal V}_2=
\{0, 1, -1, 0, -1, 1 \}^T,\;\;S=1/2,
\label{v4_2}
\end{equation}
\begin{equation}
{\cal V}_3=
\{ 2, -1, -1, -2, 1, 1 \}^T,\;\;S=1/2,
\label{v4_3}
\end{equation}
\begin{equation}
{\cal V}_4=
\{-2, 1, 1, -2, 1, 1 \}^T,\;\;S=1/2,
\label{v4_4}
\end{equation}
\begin{equation}
{\cal V}_5=
\{1, 1, 1, 1, 1, 1 \}^T,\;\; S=3/2,
\label{v4_5}
\end{equation}
\begin{equation}
{\cal V}_6=
\{-1, -1, -1, 1, 1, 1 \}^T,\;\; S=3/2.
\label{v4_6}
\end{equation}

\end{document}